\documentclass[12pt,a4paper]{article}
\usepackage[top=30mm,bottom=40mm,left=25mm,right=25mm,footskip=20mm]{geometry}
\usepackage{epsf}
\usepackage{amsmath,amsfonts,amssymb}
\usepackage{graphicx,color}
\usepackage{indentfirst}
\usepackage{ulem}

\newcommand{\nc}{\newcommand}
\nc{\rnc}{\renewcommand}
\nc{\nn}{\nonumber}
\nc{\db}{\displaybreak[0]\\}
\nc{\ds}{\displaystyle}
\rnc{\c}[1]{\textcolor{blue}{#1}}
\rnc{\[}{\left[}
\rnc{\]}{\right]}
\rnc{\(}{\left(}
\rnc{\)}{\right)}
\nc{\lt}{\left\{}
\nc{\rt}{\right\}}
\rnc{\o}{\omega}
\rnc{\a}{\alpha}
\rnc{\b}{\beta}
\nc{\lam}{\lambda}
\nc{\sig}{\sigma}
\nc{\eps}{\epsilon}
\nc{\vp}{\varphi}
\nc{\del}{\delta}
\nc{\D}{\Delta}
\rnc{\th}{\theta}
\nc{\Th}{\Theta}
\nc{\z}{\zeta}
\nc{\g}{\gamma}
\nc{\ch}{\cosh}
\nc{\sh}{\sinh}
\nc{\bra}{\langle}
\nc{\ket}{\rangle}
\rnc{\i}{{\rm i}}
\rnc{\d}{{\rm d}}
\nc{\Res}[2]{{\rm Res}\[\left.#1\right|_{#2}\]}
\nc{\intall}{\int_{-\infty}^\infty}
\nc{\inth}{\int_0^{\infty}}
\nc{\atan}{\tan^{-1}}
\nc{\sign}{\mbox{sign}}
\nc{\xxz}{H_{\text{XXZ}}}
\nc{\sm}[2]{\sum_{#1}^{#2}}
\nc{\tr}[1]{\mbox{tr}_{#1}}
\nc{\til}{\tilde}
\nc{\hh}{\check{h}}
\nc{\RR}{\check{R}}
\nc{\dlam}{\frac d{d\lam}}
\nc{\cd}{\cdots}
\nc{\re}{\eqref}
\nc{\fr}[2]{\frac{#1}{#2}}

\nc{\obib}[2]{\frac{d {#1}}{d {#2}}}
\nc{\pbib}[2]{\frac{\partial {#1}}{\partial {#2}}}
\rnc{\k}{\kappa}

\begin{document}

\title{Finite-temperature behavior of an impurity in the spin-1/2 XXZ chain }

\author{Ryoko Yahagi\footnote{e-mail: yahagi@hep.phys.ocha.ac.jp},
Jun Sato$^{\S}$\footnote{e-mail: jsato@jamology.rcast.u-tokyo.ac.jp}, 
and Tetsuo Deguchi\footnote{e-mail: deguchi@phys.ocha.ac.jp} 
}
\date{\today}
\maketitle
\begin{center} 
Department of Physics, Graduate School of Humanities and Sciences, \\
Ochanomizu University \\
2-1-1 Ohtsuka, Bunkyo-ku, Tokyo 112-8610, Japan 
\end{center} 

\begin{center}  
$^\S$ Research Center for Advanced Science and Technology, University of Tokyo, \\ 
4-6-1 Komaba, Meguro-ku, Tokyo 153-8904, Japan
\end{center}

\begin{abstract}
We study the zero- and the finite-temperature behavior of the integrable spin-1/2 XXZ periodic chain with an impurity by the algebraic and thermal Bethe ansatz methods. We evaluate the impurity local magnetization at zero temperature analytically and derive the impurity susceptibility exactly from it. In the graphs of the impurity specific heat versus temperature, we show how the impurity spin becomes more liberated from the bulk many-body effect as the exchange coupling between the impurity spin and other spins decreases, and also that in low temperature it couples strongly to them such as the Kondo effect.  Thus, we observe not only the crossover behavior from the high- to the low-temperature regime but also another one from the $N$-site chain to the $(N-1)$-site chain with a free impurity spin.  We also show that the estimate of the Wilson ratio at a given low temperature is independent of the impurity parameter if its absolute value is small enough with respect to the temperature, and the universality class is described by the XXZ anisotropy in terms of the dressed charge. 
\end{abstract}

\newpage 
%
%
\setcounter{equation}{0}
\setcounter{figure}{0}
\section{Introduction}\label{int}

Several decades have passed since the Kondo effect  was discovered experimentally  
in the 1930s \cite{H} and was first explained theoretically in the 1960s \cite{K64}. 
However,  the Kondo problem and related subjects are still quite attractive 
in both theoretical and experimental studies \cite{Nozieres,Hewson,Kondo}.  
Quantum impurity systems show universal critical behavior at low  temperature, 
which is characterized by the ratio of the impurity magnetic susceptibility $\chi_{imp}$ 
to the impurity specific heat $c_{imp}$ with temperature $T$  \cite{Wi}
\begin{eqnarray} 
r =  \fr{\pi^2}{3}  \fr{\chi_{imp}} {c_{imp}/T} . \label{eq:W-ratio}
\end{eqnarray}
We call it the Wilson ratio.  It was exactly shown by the Bethe ansatz  
that it is given by 2 for the Kondo model \cite{AFL,TW}.

The effect of an impurity embedded in a one-dimensional interacting quantum system 
or  in a Tomonaga-Luttinger liquid \cite{TL} has been one of the most attractive topics 
during the 1990s and also in the last decade. It has been 
investigated in a variety of systems  by different 
methods such as renormalization group techniques \cite{Kane,Furusaki,Matveev}, 
conformal field theories (CFT) \cite{AL,Frojdh}, 
numerical techniques with CFT \cite{Eggert-Affleck,Oshikawa}, and 
the Bethe-ansatz method 
\cite{AJ,Schlottmann1991,Sacramento,Eckle,FZ1997,KZ,Sch,ZK2000,Bortz,Zbook}.    
Recently, it is studied by functional renormalization group \cite{FRG} and also experimentally in a quasi-one-dimensional conductor \cite{thermionic}.   
However, it is still rare that the finite-temperature thermodynamic behavior is explicitly and exactly shown by a theoretical method for a large but finite lattice system without making any approximation or assumption.


In this paper we show explicitly the finite-temperature behavior of an integrable model of the spin-1/2 XXZ periodic chain with a spin-1/2 impurity.  We remark that the spin part of the Kondo model is quite similar to the Heisenberg spin chain, i.e. the XXX model. It is thus interesting to study the integrable spin-chains with the spin-1/2 impurity which are solved by the Bethe ansatz. In the XXZ impurity model we  investigate exactly the effect of an impurity embedded in the interacting quantum spin chain similarly as the Kondo model.  Schlottmann studied the integrable spin-$S$ XXZ spin chain with one spin-$S'$ impurity \cite{Sch} in association with the multi-channel Kondo effect \cite{Schlottmann1991,Sacramento,Bortz}. 
The spin-1/2 impurity in the open spin-1/2 XXX chain was studied by Frahm and Zvyagin \cite{FZ1997} and it was shown that the Kondo-like temperature exists,  while the periodic XXZ chain with an impurity was studied by Eckle et al. \cite{Eckle}.    
However, the analytic expressions of the impurity magnetization and the impurity specific heat at low temperature in Ref. \cite{Sch} are not accurate enough to evaluate the Wilson ratio correctly. They are not consistent with the numerical estimates evaluated by solving the Bethe ansatz equations, in particular, when the impurity parameter is small.   
Moreover, the crossover temperature has not been  evaluated numerically, while its analytic expression \cite{Sch} is not appropriate. 
The thermodynamic behavior of the spin-1/2 impurities in the spin-1/2 XXZ chain 
was shown in Ref. \cite{ZK2000} for several distributions of the impurity parameters, and in the homogeneous case  the results are consistent with ours. 
In the present paper we show how the thermodynamic behavior of the single spin-1/2 impurity depends on the impurity parameter. Although it looks simpler than that of many impurities,  it is still nontrivial and interesting from the viewpoint of universality and
the crossover from the $N$-site to the $(N-1)$-site chain.

We consider the integrable model of the spin-1/2 XXZ periodic chain with a spin-1/2 impurity,  i.e.,  the case of  $S=S'=\fr12$ in the XXZ  impurity model of Ref. \cite{Sch}.  We investigate the finite-temperature behavior of the integrable XXZ impurity model by numerically solving the truncated integral equations of the thermal Bethe ansatz \cite{TS,T73,FZ,T}, and evaluate the specific heat and entropy numerically. By plotting graphs of the impurity specific heat and entropy versus temperature we show that the impurity spin gradually becomes a free spin through pseudo-decoupling from other spins,  while in low temperature it couples strongly to them such as the Kondo effect. Here,  we call it {\it pseudo-decoupling},  if the exchange coupling between the impurity spin and other spins is very small but nonzero. It occurs in the present model when the absolute value of the impurity parameter is very large.    
We then show that the estimate of the Wilson ratio at a given low temperature is independent of the impurity parameter if the absolute value of the impurity parameter is small with respect to the given temperature.

We evaluate the local magnetization at the impurity site under a given small magnetic field at zero temperature and derive its analytic expression by the two methods: by identifying the impurity contribution in the Bethe ansatz equations physically; by calculating it directly through the algebraic Bethe ansatz method \cite{Kitanine2000}. We show that they are consistent: the impurity contribution identified with the Bethe ansatz equations is valid. 

Let us denote the Hamiltonian of the XXZ spin chain with the spin-1/2 impurity consisting of $N$ sites by ${\cal H}_N(x)$, which depends on the impurity parameter,  $x$. If the absolute value of the impurity parameter, $|x|$, becomes large, the exchange coupling of the impurity spin to neighboring spins becomes small, and hence the Hamiltonian of the $N$-site chain is decomposed into that of the $(N-1)$- site chain,  ${\cal H}_{N-1}(x)$, and that of the impurity site, ${\cal H}_{imp}$, as follows.    
\begin{equation} {\cal H}_N(x) =  {\cal H}_{N-1}(0) + {\cal H}_{imp} + O(e^{-|x|}) \qquad \mbox{for} \quad |x| \gg 1    \, . \label{eq:decomposedH} \end{equation}
As we shall see in \S 2  the term ${\cal H}_{N-1}(0)$ gives the Hamiltonian of the spin-1/2 periodic XXZ chain with no impurity. Applying a small magnetic field, $h$, on the whole system of $M$ down spins we have $M < N/2$.  Here the Hamiltonian is given by  ${\cal H}^{'}={\cal H}_N(x)- 2 h S^z$, where $S^z$ denotes the eigenvalue of the $z$-component of the total spin operator. We also assume that the XXZ spin chain, ${\cal H}_{N-1}(0)$, has the anti-ferromagnetic interactions. Then,  any down spin on the impurity site is attracted to the the $(N-1)$-site chain due to the anti-ferromagnetic interactions, since the ground-state energy of ${\cal H}_{N-1}(0)$ with $M$ down spins decreases as the number of down spins, $M$, increases. 
We thus suggest that the impurity spin tends to be given only by an up-spin and all the $M$ down-spins on the whole chain tend to be located on the $(N-1)$-site chain 
under non-zero magnetic field if the absolute value of the impurity parameter is large, i.e., in the case of pseudo-decoupling. Moreover, if the impurity spin is completely decoupled from other spins, it becomes a free spin. 
We thus have the crossover from the $N$-site chain to the $(N-1)$-site chain with a free impurity spin.

In the XXX case of the integrable impurity model the Wilson ratio has been evaluated numerically and systematically by applying the quantum transfer-matrix method \cite{KZ}, in which it is necessary to consider logarithmic corrections for the magnetization in low temperature due to the $SU(2)$ symmetry \cite{EAT}. In the XXZ impurity model, however,  it is not  necessary to consider logarithmic corrections, and hence it is easier to derive the susceptibility at a very low temperature.


The contents of the present paper consist of the following. In section 2, we introduce the spin-1/2 XXZ chain with a spin-1/2 impurity and give an explicit expression of the Hamiltonian in terms of the spin operators. We show that the impurity spin operator becomes decoupled  from the other spin operators if we send the absolute value of the impurity parameter to infinity.  In section 3 we derive an analytic solution to the ground state under a weak magnetic field by the Wiener-Hopf method and present an analytic expression of the impurity susceptibility at zero temperature. We then calculate the local magnetization on the impurity site systematically by the algebraic Bethe ansatz. We  confirm the analytical result of the impurity magnetization with numerical plots. 
Furthermore, we  evaluate the impurity susceptibility at zero temperature numerically by solving the Bethe ansatz equations, and show that the numerical estimates are consistent with the analytic expression if we take the system size large enough.  
In section \ref{spe} we evaluate the impurity specific heat numerically by solving the truncated integral equations of the thermal Bethe ansatz. 
In the plots of the impurity specific heat we observe that 
the impurity spin tends to become a free spin as the absolute value of the impurity parameter becomes very large, which is clearly seen in high temperature,  
while it couples strongly to other spins in low temperature, that is, the Kondo effect appears.   
We give an analytic expression of the impurity specific heat at a low temperature, which is valid if the absolute value of the impurity parameter is small with respect to the given low temperature. 
We numerically determine the crossover temperature from the high- to low-temperature regimes as a function of the impurity parameter and give a good fitting formula to the data. 
In the last section, we derive the universal value of the Wilson ratio at low temperature 
and express it in terms of the dressed charge of the XXZ spin chain \cite{BIK1986}. We show that the estimate of the Wilson ratio gives the universal value if 
  the absolute value of the impurity parameter is small enough 
with respect to the temperature.

%
%
\setcounter{equation}{0}
\setcounter{figure}{0}
\section{Integrable model of an impurity}\label{mo}

\subsection{An inhomogeneous transfer matrix of the XXZ spin chain}

We now introduce one of the integrable XXZ models with 
a spin-1/2 impurity through the algebraic Bethe ansatz.
Let $V_0, V_1, \ldots , V_N$ be two-dimensional vector spaces over complex numbers ${\bf C}$. Here we denote by  $N$  the number of sites on the XXZ spin chain.
We call $V_0$  the auxiliary space and the tensor product 
$V_1 \otimes \cdots \otimes V_N$ the quantum space, 
where $V_j$ corresponds to the $j$th site of the XXZ spin chain 
for $j=1, 2, \ldots, N$. 
On the tensor product $V_0 \otimes V_n$ 
we  define the $R$-matrix by  
\begin{equation}
  R_{0n}(\lam)=
\begin{pmatrix}
1&&& \\
&b(\lam)&c(\lam)& \\
&c(\lam)&b(\lam)& \\
&&&1
\end{pmatrix}_{\!\![0,n]}  . 
\end{equation}
Here the suffix $[0, n]$ means that the matrix is defined for the 
tensor product $V_0 \otimes V_n$.  
We define the transfer matrix of the spin-1/2 XXZ spin chain 
by the trace of the $N$th product of the $R$ matrices 
$R_{0j}(\lam-\xi_j)$ for $j=1, 2, \ldots, N$, over the auxiliary space 0: 
\begin{equation}
 \tau_{1\cdots N}(\lam|\xi_1,\cdots,\xi_N)
 = \mbox{tr}_0 R_{0N}(\lam-\xi_N) \cdots R_{01}(\lam-\xi_1), 
\end{equation}
where $\xi_j$ ($j=1, 2, \ldots, N)$ are arbitrary. 
We call them the inhomogeneity parameters.  
The functions $b(\lam)$ and $c(\lam)$ are given by 
\begin{eqnarray}
 b(\lam)=\frac{\vp(\lam)}{\vp(\lam+i\zeta)}, \quad  
c(\lam)=\frac{\vp(i\zeta)}{\vp(\lam+i\zeta)} \, ,  
\end{eqnarray}
where  $\vp(\lam)=\sinh\lam$ and 
$\Delta=\cos \zeta$ with $0 < \zeta \le \pi$ in the gapless regime of the XXZ spin chain. 
We  consider only the gapless regime throughout the paper,.

%
%
\subsection{Definition of the XXZ chain with an impurity: $\xxz(x)$ }

We now define the Hamiltonian of the impurity model, $\xxz(x)$. 
We assume that only the 1st site has a different value 
for the  inhomogeneity parameter $\xi_1$: 
\begin{equation} 
\xi_1 =\frac{i\zeta}{2}-x  \, ,  \quad \mbox{for } \quad x \in \mathbb{R} \, ,   
\end{equation}
while the inhomogeneous parameters of the other sites $\xi_j$ for $j=2, 3, \ldots, N$,  
have the same value $i \zeta/2$. 
Here we assume that the impurity parameter $x$ is real.  
We  define  the Hamiltonian of the impurity model, $\xxz(x)$, 
by the logarithmic derivative of the transfer matrix of the spin-1/2 XXZ chain  
as follws. 
\begin{eqnarray}
\label{h}
 \begin{split}
\xxz(x)&=\frac{\vp(i\zeta)}2\dlam\log\tau_{1\cdots N}\left(\lam\Big|\xi_1,\frac{i\zeta}{2},\cdots,\frac{i\zeta}{2}\right)\Big|_{\lam\to\frac{i\zeta}{2}} 
 \end{split} . 
\end{eqnarray}
By putting  $x=0$ (i.e., $x \to 0$), the Hamiltonian $\xxz(x)$  reduces to the standard homogeneous spin-1/2 XXZ Hamiltonian, which has no impurity. It is clear that the Hamiltonian (\ref{h}) is integrable. It is a consequence of the construction in terms of  the algebraic Bethe ansatz.

Let $M$ be the number of down spins. 
The Bethe-ansatz equations (BAE) for the impurity Hamiltonian (\ref{h}) 
are given by 
\begin{equation}
(N-1) \theta_1(z_l) + \theta_1(z_l+x) =   
2\pi I_l +\sum^M_{j\neq l}\theta_2(z_l-z_j) ,  \quad 
\mbox{for} \, \,   \l=1,\ldots, M , \label{bae}  
\end{equation}
where $I_l$ are called the Bethe quantum numbers, and they are 
integers or half-integers according to the following rule:  
\begin{equation}
 I_l \equiv \frac{N-M+1}{2} \pmod{1}, \ \ \mbox{for} \, \,  l=1, 2, \ldots, M . 
\end{equation}
Here the functions $\theta_n(z)$ are given by 
\begin{equation}
 \theta_n(z) =  i \log \left[-\frac{\vp(z+\frac{in\z}{2})}{\vp(z-\frac{in\z}{2})}\right] \, . 
\end{equation}
We remark that in the special case, i.e. for the XXX model with $x=1$, 
the BAEs (\ref{bae}) are similar to the Kondo model. 

By solving the Bethe-ansatz equations (\ref{bae}), we can evaluate 
the energy of the Bethe ansatz eigenvector with $M$ down spins. 
Let $z_{\ell}$ be a solution of the Bethe ansatz equations  (\ref{bae}).  
The energy for the eigenstate with Bethe roots 
$z_{\ell} $ is expressed in terms of the set of Bethe roots $z_i$ as  
\begin{eqnarray}\label{en}
 E= -\ds{\sum^M_{l=1}\fr{ \sin^2 \z}{\cosh 2z_l-\cos\z}}\, .  
 \end{eqnarray}
We remark that the energy depends on the impurity parameter $x$,  
since the solutions $z_j$ contain the information of $x$, though it does not appear explicitly.

%
%
\subsection{Expression of $\xxz(x)$ in terms of local spin operators} 

We now present an explicit expression of the XXZ Hamiltonian with a spin-1/2 impurity, 
$\xxz(x)$, defined by eq. \re{h} 
in terms of the local spin operators. 
The expression is useful for confirming analytic results. Making use of the explicit expression 
of $\xxz(x)$ we can perform the exact diagonalization of the Hamiltonian \re{h}, 
and compare numerical results with analytic or numerical results which are obtained 
by solving the Bethe ansatz equations \re{bae}. 

Through a direct but straightforward calculation we obtain the following compact expression of $\xxz(x)$ in terms of local spin operators $S_j^{\pm}$ and $S^z_j$. 
\begin{eqnarray}\label{eq:hxxz} 
 \xxz(x)
&=&\sum_{n=2}^{N-1}  \left[ \frac12(S_n^+S_{n+1}^-+S_n^-S_{n+1}^+)
+\Delta\left(S_n^zS_{n+1}^z-\frac14\right)\right] \nn \\
&+&c^+c^-
\left[ \frac{\vp'(x)}2(S_N^+S_{1}^-+S_N^-S_{1}^+)
+\Delta\left(S_N^zS_{1}^z-\frac14\right)\right] \nn \\
&+&c^+c^-\left[
\frac{\vp'(x)}2(S_1^+S_2^-+S_1^-S_2^+)+\Delta S_1^zS_2^z\right] \nn \\
&+&b^+b^-\left[
\frac{\D}2(S_N^+S_2^-+S_N^-S_2^+)+\Delta S_N^zS_2^z\right] 
-\fr{\Delta}{4} \nn \\
&+&b^+c^-\Biggl[
 \vp'(x)   (S_N^+S_2^--S_N^-S_2^+)S_1^z \nn \\
 &&\ \ \ \ \ \ \ \ -\D(S_N^+S_1^--S_N^-S_1^+)S_2^z \nn \\
 &&\ \ \ \ \ \ \ \ -\D(S_1^+S_2^--S_1^-S_2^+)S_N^z \Biggr], 
\end{eqnarray}
where the symbols $b^\pm$ and $c^\pm$ are given by  
\begin{eqnarray}\label{bc}
 b^\pm=b(\pm x)=\frac{\vp(x)}{\vp(x\pm i\zeta)},
 \ \ \ \ 
 c^\pm=c(\pm x)=\pm\frac{\vp(i\zeta)}{\vp(x\pm i\zeta)}.
\end{eqnarray}
Here, the spin operators $S_j^{\pm}$ and $S_j^z$ are expressed in terms of the 
Pauli matrices  as follows. 
\begin{equation} 
S_j^{\pm} = {\frac 1 2} \sigma_j^x \pm i {\frac 1 2} \sigma_j^y , \quad 
 S_j^z = {\frac 1 2 } \sigma_j^z \, ,  \quad  \mbox{for} \quad j = 1, 2, \ldots, N \, .  
\end{equation}  
Some details of the derivation of \re{eq:hxxz} are given in Appendix \ref{apph}.

We remark that the XXZ Hamiltonian with the spin-1/2 impurity, $\xxz(x)$, given in eq. \re{eq:hxxz}   
is Hermitian when the impurity parameter $x$ is real. 

In the cases of $N \leqq 10$ we have confirmed explicitly that the lowest energy level of the XXZ Hamiltonian with the spin-1/2 impurity \re{eq:hxxz} is  completely 
consistent with the eigenvalue evaluated through eq. \re{en} 
from the ground-state solution of the Bethe ansatz equations. 
We therefore  assume that the quantum numbers $I_{\ell}$ of the ground state do not change  under the existence of the impurity. They are given by 
\begin{equation} 
I_{\ell}= j - {\frac {N-M-1} 2} \, \, \quad  \mbox{for} \, \,  j=1 ,2, \ldots, M \, .  
\end{equation} 
We can confirm this assumption by directly diagonalizing the XXZ Hamiltonian with an impurity for a larger number of sites.

We should remark that the integrable model of an impurity given by eq. \re{h} corresponds to the spin-$\fr12$ case of the integrable XXZ model with the spin-1/2 impurity studied by Schlottmann \cite{Sch}. 
Here we recall that the spin part of the Kondo model \cite{Kondo} is similar to  
the spin-1/2 XXX chain. 

%
%
\subsection{Pseudo-decoupling of the impurity spin for large $|x|$}

If one takes the limit $x\to\infty$, the Hamiltonian  \re{eq:hxxz}  reduces to 
\begin{eqnarray}\label{hinf}
 \xxz(\infty)
 &=&\sm{n=2}{N-1}  \Bigl[ \frac12(S_n^+S_{n+1}^-+S_n^-S_{n+1}^+)
 +\D\(S_n^zS_{n+1}^z-\frac14\)\Bigr] \nn \\
 && +\frac\D2\(S^+_NS^-_2+S^-_NS^+_2\)+\D\(S_N^zS_2^z-\frac14\) \nn \\
 && +2\sqrt{1-\D^2}\(S^x_NS^y_2-S^y_NS^x_2\)S_1^z. 
\end{eqnarray}
The impurity spin, i.e., the spin operators defined on the 1st site such as $S_1^{\pm}$ and $S_1^z$,  appears only in the last term of the reduced Hamiltonian  in eq. (\ref{hinf}).  
The exchange coupling of the impurity spin to the neighboring spins vanishes in  $\xxz(\infty)$. In terms of eq. \re{eq:decomposedH} the bulk part ${\cal H}_{N-1}(0)$ and the impurity part ${\cal H}_{imp}(0)$ corresponds to the first two lines and the third line of the right-hand side of \re{hinf}, respectively.

If one takes the XXX limit ($\D\to1$), the terms associated with the impurity site vanish.
The Hamiltonian now reduces to the homogeneous spin-1/2 XXX Hamiltonian defined on the $N-1$ sites 
from site 2 to site $N$:
\begin{eqnarray}\label{xxxinf}
H_{\rm XXX}(\infty) = 
 \sm{n=2}N \left[ S_n^xS_{n+1}^x+S_n^yS_{n+1}^y+S_n^zS_{n+1}^z-\fr14\right].
\end{eqnarray}

Under a magnetic field, $h>0$, the number of down spins, $M$, becomes smaller than that of the half-filling case, i.e., $M < N/2$.  Let us assume that when 
the impurity spin is pseudo-decoupled from other spins, we can effectively define  
the energy of the bulk part ${\cal H}_{N-1}(0)$ for the eigenstates with $M$ down spins in the bulk chain and denote it by $E_{N-1}^{bulk}(M)$.  
We then suggest that any down spin on the impurity site will be attracted to the bulk chain since the bulk energy decreases as the number of down spins in the bulk increases: $E_{N-1}^{bulk}(M) < E_{N-1}^{bulk}(M-1)$ due to the anti-ferromagnetic interactions. 
Here the total energy reduction including the Zeeman term is of the order of $1/N$. 
We also remark that the magnetic energy term, $- 2 h S^z$, in the total Hamiltonian  ${\cal H}^{'}$ does not change whether the impurity spin has an up spin or down spin in the sector of $M$ down spins.  

We thus suggest that the impurity spin tends to be  given by an up spin and all the $M$ down-spins on the whole chain tend to be located on the $(N-1)$-site chain for any number $M$ of down spins if pseudo-decoupling occurs under a magnetic field. Furthermore,  it is consistent with another viewpoint 
that the impurity spin becomes a free spin  if it is decoupled with other spins completely.  
We therefore have the crossover from the $N$-site chain to the $(N-1)$-site chain 
with the free impurity spin as the exchange coupling between the impurity spin and other spins decreases to zero.

In the graphs of the impurity specific heat in \S 4.  we show how the impurity spin becomes free from the other spins in the case of  large absolute values of $x$

%
%
\setcounter{equation}{0}
\setcounter{figure}{0}
\section{Magnetic susceptibility due to impurity}\label{mag}

\subsection{The Bethe anzatz equations with an impurity}

We shall calculate the magnetic susceptibility by adding a small magnetic field $h$ to the Hamiltonian defined by \re{h}:
\begin{eqnarray}\label{hh}
H^{'}=\xxz(x)-2h\sum_{n=1}^N S_n^z.
\end{eqnarray}
We recall that the Hamiltonian $\xxz(x)$ is expressed explicitly in terms of spin operators 
in \re{eq:hxxz}. 

Let us assume that the Bethe quantum numbers $I_{\ell}$ 
of the ground state do not change  under the existence of the impurity.  
By taking the thermodynamic limit of the BAE \re{bae}
through the Euler-Maclaurin formula,  
we derive the following inregral equation: 
\begin{eqnarray}\label{baeh}
 \rho(z, x) + \int^B_{-B}a_2(z-z')\rho(z', x)dz'
 =\frac{1}{N} \lt (N-1)a_1(z)+a_1(z+x) \rt.
\end{eqnarray}
Here parameter $B$ denotes the Fermi point determined by the magnetic field $h$, 
and $\rho(z, x)$  the density of the Bethe roots in the ground-state solution 
of the BAE (\ref{bae}), 
where functions $a_n(z)$ for $n=1, 2, \ldots, $ are defined by 
\begin{eqnarray}
 a_n(z)=
 \frac{\theta_n^{'}(z)}{2\pi} =-\frac{i}{2\pi}
 \frac{\vp(in\z)}{\vp(z+\frac{in\z}{2})\vp(z-\frac{in\z}{2})}. \label{eq:an}
\end{eqnarray}
We remark that  we have the infinite Fermi point, $B=\infty$, for $h=0$.  

We define the Fourier transform of a given function $f(z)$ by 
\begin{equation} 
\tilde{f}(k)=\int^{\infty}_{-\infty} f(z) e^{ikz} dz . 
\end{equation} 
Let us denote  by $\rho_0(z, x)$ the root density for the ground-state solution of the BAE 
 with no magnetic field for \re{h}. We now derive the analytic expression of  $\rho_0(z, x)$.  
By taking the Fourier transform of the following integral equation 
\begin{eqnarray}\label{bae0}
 \rho_0(z, x) + \int^{\infty}_{-\infty}a_2(z-z')\rho_0(z', x)dz'
 =\frac{1}{N} \lt (N-1)a_1(z)+a_1(z+x) \rt \, , 
\end{eqnarray}
we have 
\begin{eqnarray}\label{tilrho}
 \tilde{\rho}_0(k, x)=\fr{(N-1+e^{-ikx})\tilde{a}_1(k)}{N(1+\tilde{a}_2(k))}.
\end{eqnarray}
Here  $\til{a}_1(k)$ and $\til{a}_2(k)$ are given by  
\begin{eqnarray}\label{a}
 \til{a}_1(k)=
   \fr{\sinh \fr{k}{2} \( \pi-\z \) }
  {\sinh \fr{k\pi}{2}},
 \ \ \ \ \ 
 \til{a}_2(k)=
   \fr{\sinh \fr{k}{2} \( \pi - 2\z \) }
  {\sinh \fr{k\pi}{2}}.
\end{eqnarray}
Taking the inverse transform of \re{tilrho}, one finds
\begin{eqnarray}
 \rho_0(z, x)=
  \frac{1}{2N\z}
  \left\{(N-1)\mathrm{sech}\fr{\pi z}{\z}
  +\mathrm{sech}\fr{\pi(z+x)}{\z} \right\}.
\end{eqnarray}
In the case of $x=0$, it reduces to the bulk density, $\sig_0(z)$, which is given by
\begin{eqnarray}
 \sig_0(z)=
  \fr{1}{2\z}\mathrm{sech}\fr{\pi z}{\z}.
\end{eqnarray}
\begin{figure}[h]
 \centering
 \includegraphics[width=7cm]{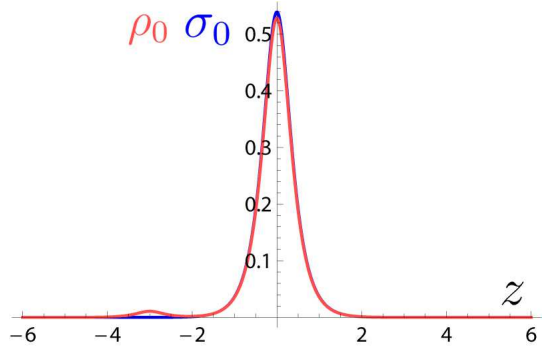}
 \caption{$\vtop{
\hbox{Root densities $\rho_0(z, x)$ and $\sigma_0(z)$ of the solution } 
\hbox{of the Bethe-ansatz equations for $N=50$ and $\D=0.6$.} 
\hbox{$\rho_0(z, x)$ ($x=3$) with an impurity (red curve); }
\hbox{$\sig_0(z)$ ($x=0$) with no impurity (blue curve). }
\hbox{Note that the red curve has a small peak at $z=-x=-3$. }
}$}
 \label{frho}
\end{figure}
Hereafter we shall sometimes abbreviate $\rho(z,x)$ and $\rho_0(z, x)$    
as  $\rho(z)$ and $\rho_0(z)$, respectively. 

Let us evaluate the largest value of the impurity parameter $x$ for which  thermodynamic expressions are valid.  We shall compare several numerical data obtained for the system of a finite number of $N$ with such analytic expressions that are derived in the thermodynamic limit. We approximately estimate it by setting $\sigma_0(x) > 1/N$. If the root density is smaller than $1/N$, then the continuous approximation to  
the discrete distribution of the Bethe roots should be not good. 
From Fig. \ref{frho} we have $x=1.2$ for $N=50$.  
  
Let us introduce a function $R(z)$ by 
\begin{eqnarray}
 \int^{\infty}_{-\infty}\(a_2(z-z')R(z-z'')\)dz=a_2(z''-z')-R(z''-z') 
\end{eqnarray}
or equivalently by 
\begin{eqnarray}
 \tilde{a}_2(k)\tilde{R}(k)=\tilde{a}_2(k)-\tilde{R}(k) . 
\end{eqnarray}
Multiplying $R(z-z'')$, the integral equation \re{baeh} and  \re{bae0} are merged into the following: 
\begin{eqnarray}\label{rho}
 \rho(z)=\rho_0(z)+\int_{|z'|>B} R(z-z')\rho(z')dz' . \label{eq:fund} 
\end{eqnarray}
In the Wiener-Hopf method the integral equation \re{eq:fund} plays a fundamental role.   

We denote  by $s^z(x)$ the magnetization per site under magnetic field $h$.  
We shall give the relation between the Fermi point $B$ and the magnetic field $h$ by the 
Wiener-Hopf method in \S 3.2. 
Let us recall that the energy eigenvalue of the Hamiltonian $\xxz(x)$ is given by  
\begin{equation}
{\frac E N}  = - \pi \sin \zeta \, \, {\frac 1 N} \sum_{j=1}^{M} a_1(z_j) \, . 
\end{equation}
We consider the difference of the ground-state energy  per site under magnetic field $h$ 
from that of no magnetic field, and we denote it  by $e(x)$.
Then, physical quantities such as $s^z(x)$  and  $e(x)$ 
are written in terms of the inetgrals of the root density $\rho(z)$ as follows. 
\begin{eqnarray}
 s^z(x) &=&\fr12-\int_{|z|<B} \rho(z) dz
 = \fr{\pi}{2\pi-2\z}\int_{|z|>B} \rho(z) dz, \\
 e(x) &=&-\pi\sin\z \int_{|z|<B} a_1(z) \rho(z) dz 
 -\( -\pi\sin\z \intall a_1(z) \rho_0(z) dz \) \nn \\
 &=& \pi \sin \z \int_{|z|>B}\sigma_0(z)\rho(z)dz.
\end{eqnarray}
Hereafter, we shall often abbreviate the magnetization per site and the energy per site 
simply as the magnetization and the energy, respectively.

%
%
\subsection{Wiener-Hopf  integral equation with an impurity}
 
Let us define function $y(z,x)$ by 
\begin{equation}
 y(z,x) =\rho(z+B, x) . 
\end{equation}
By making use of the invariance of eq. \re{baeh} under the replacement of $z$ and $x$ by $-z$ and $-x$, respectively,  we show the following relation:  
\begin{equation}
 y(z, -x) =\rho(-z-B, x) . 
\end{equation}
We define functions $\breve{y}_{\pm}(k, x)$ by 
\begin{equation} 
 \breve{y}_{\pm}(k,x) = \intall \mathcal{H}(\pm z)y(z,x)e^{ikz}dz . 
\end{equation}
Here $\mathcal{H}(z)$ is the Heaviside step function: $\mathcal{H}(z)=1$ for $z > 0$ and 
$\mathcal{H}(z)=0$ for $z < 0$. 
In terms of functions $y(z, \pm x)$ the integral equation \re{rho} is now expressed as follows. 
\begin{eqnarray}
 y(z,x) &=&\rho_0(z+B)+\inth R(z-z')y(z',x)dz' \nn \\
 && +\inth R(z+z'+2B)y(z',-x)dz', \label{eq:WHE1} \\ 
 y(z,-x) &=&\rho_0(-z-B)+ \inth R(z-z')y(z',-x)dz' \nn \\
 && +\inth R(z+z'+2B)y(z',x)dz'. \label{eq:WHE2} 
\end{eqnarray}

Let us now assume that the Fermi point $B$ is very large. 
The magnetization and the energy are expressed in terms of  
functions $\breve{y}_{\pm}(k, x)$  as follows. 
\begin{eqnarray}
 s^z(x) &=&\fr{\pi}{2\pi-2\z} \lt \breve{y}_+(0,x)+\breve{y}_+(0,-x) \rt, \\
 e(x) &\simeq&\fr{\pi \sin \z}{\z} e^{-\fr{\pi B}{\z}}
 \lt \breve{y}_+(\fr{i\pi}{\z},x)+\breve{y}_+(\fr{i\pi}{\z},-x)\rt. \label{eq:energy-yp}
\end{eqnarray}
Here we recall that 
eq. (\ref{eq:energy-yp}) is valid only for large values of $B$: $\exp(-\pi B/\zeta) \ll 1$.

It is known that integral equations such as  eqs. \re{eq:WHE1} and \re{eq:WHE2} 
can be solved by the Wiener-Hopf method (see \cite{Kondo},  \cite{YY}, \cite{T}). 
 In the following analysis we neglect 
the terms with $R(z+z'+2B)$ in eqs. \re{eq:WHE2} 
since the corrections from them are negligible for $h\ll 1$. 
 Expressing $k_n= \pi i(2n+1)/\zeta$ for $n \in \mathbb{Z}$, 
we thus have 
\begin{eqnarray}
 \breve{y}_+(k,x) = 
 \fr{i}{N\z}G_+\(\fr{\z k}{2}\)\sum_{n=0}^{\infty}
 \fr{(-1)^n G_-\(-\fr{\z k_n}{2}\)e^{ik_nB}}{k+k_n}\(N-1+e^{-ik_nx}\),
\end{eqnarray}
where functions $G_{\pm}(k)$ are given by 
\begin{equation}
 G_+(i\pi z)
 =\sqrt{2\pi(1-\fr{1}{\g})}\(\fr{(\g-1)^{\g-1}}{\g^{\g}}\)^z
 \fr{\Gamma(\g z+1)}{\Gamma(\fr12+z)\Gamma((\g-1)z+1)}, 
\end{equation}  
with  $\g={\pi}/{\z}$,  and by the following relation:  
\begin{equation} 
 G_-(z)=G_+(-z) \, . 
\end{equation}
Then we have
\begin{eqnarray}
 s^z(x) &=&\fr{\pi}{2\pi-2\z}\fr{2i}{N\z}G_+(0)\sum_{n=0}^{\infty}
 \fr{(-1)^n G_-\(-\fr{\z k_n}{2}\)e^{ik_nB}}{k_n}\(N-1+\cosh ik_nx\), \\
 e(x) &=&\fr{\pi \sin \z}{\z} e^{-\fr{\pi B}{\z}}\fr{2i}{N\z}
 G_+\(\fr{i\pi}{2}\) \nn \\
 &&\times\sum_{n=0}^{\infty}
 \fr{(-1)^n G_-\(-\fr{\z k_n}{2}\)e^{ik_nB}}
 {\fr{i\pi}{\z}+k_n}\(N-1+\cosh ik_nx\).
\end{eqnarray}
However, the Fermi point $B$ remains unknown, yet. 
It is determined as a function of $h$ by the following condition: 
\begin{eqnarray}
 \fr{\partial}{\partial B} \left( e(x) - 2 h s^z(x) \right) = 0 .
\end{eqnarray}
We therefore have 
\begin{eqnarray}
 \exp\left( {-\fr{\pi B}{\z}} \right)
 =\fr{G_+(0) \z h}{(\pi-\z)\sin \z G_+\(\fr{i\pi}{2}\)}.  \label{eq:h-B}
\end{eqnarray}
\begin{figure}[h]
 \centering
 \includegraphics[width=7cm]{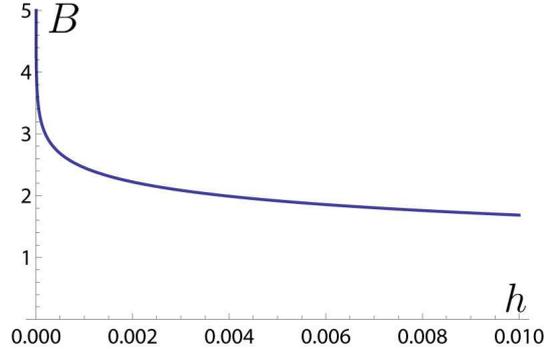}
 \caption{$\vtop{
\hbox{Fermi point $B$ as a function of  the magnetic field $h$. }
}$}
 \label{fb}
\end{figure}
The relation between the Fermi point $B$ and the magnetic field $h$ is shown explicitly 
in Fig. \ref{fb}. The Fermi point $B$ plays the role  of the cut-off in the integral of eq. \re{baeh}.  
The cut-off $B$ goes to  infinity in the limit of $h\to0$. However, 
it dramatically decreases to values such as 2 or 3, which are not very large,  
 when $h$ becomes nonzero such as $h=10^{-3}$. 

By comparing Figure \ref{fb} with the profile 
of the root density $\rho_0(z)$ shown in Fig. \ref{frho}, 
we suggest that the Wiener-Hopf method is valid only 
when the impurity parameter $x$ is small enough such as $|x| < 1$
so that the small peak at $z=-x$ in the root density $\rho_0(z)$ 
is located in the interval between  the Fermi point  $z=-B$ and the origin $z=0$.

%
%
\subsection{ Impurity susceptibility at zero temperature}

Recall that the symbol  $s^z(x)$ denotes the  magnetization per site 
in the XXZ spin chain which has the spin-1/2 impurity with impurity parameter  $x$. 
Through the Wiener-Hopf method the magnetization per site $s^z(x)$ is given by   
\begin{eqnarray}
 s^z(x)
 &\simeq&\fr{i\pi}{N\z(\pi-\z)}G_+(0)
 \fr{G_-\(-\fr{i\pi}{2}\)}{\fr{i\pi}{\z}}
 \fr{G_+(0)h\z}{(\pi-\z)\sin \z G_+\(\fr{i\pi}{2}\)}
 \(N-1+\cosh \fr{\pi x}{\z}\) \nn \\
 &=&\fr{2\z h}{N\pi(\pi-\z) \sin \z}\(N-1+\cosh\fr{\pi x}{\z}\).
\label{eq:mps}
\end{eqnarray}
Here, the magnetic field $h$ is very small, 
so that only linear terms of $h$ may appear.

Let us define the magnetic susceptibility per site, $\chi(x)$,  by 
\begin{equation}\label{eq:def-chi}
\chi(x)  = \left. {\fr {\partial  \, 2s^z(x)}{ \partial h}} \right|_{h=0} . 
\end{equation}
Consequently, we have  
\begin{equation}
\chi(x)  =  \fr{4\z}{N\pi(\pi-\z) \sin \z}\(N-1+\cosh\fr{\pi x}{\z}\).
\label{eq:suscept}
\end{equation}

We now extract the impurity susceptibility $\chi_{imp}$ induced by the spin-1/2 impurity of parameter $x$. 
We first define the impurity magnetization  $s_{imp}^z(x)$ by 
\begin{equation} 
 N s^z(x) = (N-1) s^z(0) + s_{imp}^z(x) \, . 
\end{equation}
Here,  $N s^z(x)$ gives the total magnetization of the XXZ spin chain 
with the impurity parameter $x$.  We also express it by $S_{tot}^z =N s^z(x)$.  
We define the impurity susceptibility $\chi_{imp}(x)$ by 
\begin{equation} 
\chi_{imp} = \left. \frac {\partial \, 2s^{z}_{imp}(x)} {\partial h} \right|_{h=0} .   
\end{equation} 
The impurity susceptibility is therefore given by 
\begin{equation} 
\label{eq:sc-imp}
\chi_{imp}(x)= \fr{4 \z }{\pi(\pi-\z) \sin \z} \cosh\fr{\pi x}{\z} .
\end{equation}

%
%
\subsection{Local magnetization at the impurity site}

\subsubsection{Derivation through the algebraic Bethe ansatz}

Let us now derive the local magnetization at the impurity site, $s_1^z = {\frac 1 2} \langle \sigma_1^z \rangle_{g, h}$, under a small magnetic field $h$ at zero temperature. 
We take the expectation value of the local operator $\sigma_1^z$ for the ground state under the magnetic field $h$. Here we remark that if the magnetic field is zero,  the local magnetization at the impurity site vanishes. 

We now define the monodromy matrix of the spin-1/2 XXZ spin chain 
by the $N$th product of the $R$ matrices $R_{0j}(\lam-\xi_j)$ for $j=1, 2, \ldots, N$: 
\begin{equation}
R_{0, \, 1\cdots N}(\lam |\xi_1,\cdots,\xi_N)
 = R_{0N}(\lam-\xi_N) \cdots R_{01}(\lam-\xi_1), 
\end{equation}
where $\xi_j$ ($j=1, 2, \ldots, N)$ are the inhomogeneity parameters given by 
$\xi_1= i \zeta/2 -x$ and $\xi_2= \xi_3= \cdots = \xi_N = i \zeta/2$.
We define the operator-valued matrix elements of the monodromy matrix by 
\begin{equation}  
\left(
\begin{array}{cc}
A(\lambda) & B(\lambda) \\  
C(\lambda) & D(\lambda) 
\end{array}
\right) 
= R_{0, \, 1\cdots N}(\lam |\xi_1,\cdots,\xi_N) \, . 
\end{equation}

Let us denote the ground state under a small magnetic field $h$ by $|\Psi_{g,h} \rangle$. 
We express the ground-state solution of the Bethe-ansatz equations with $M$ down spins under a magnetic field $h$  as $\lambda_b$ for $b=1, 2, \ldots, M$.  
We have    
\begin{equation} 
|\Psi_{g, h} \rangle = B(\lambda_1) \cdots B(\lambda_M) | 0 \rangle 
\end{equation}
where $|0 \rangle$ denotes the vacuum state, which has no down spin. 
The Hermitian conjugate vector is given by 
\begin{equation} 
\langle \Psi_{g, h} | = 
\left( |\Psi_{g,h} \rangle \right)^{\dagger} =  (-1)^M \,  \langle 0 |  
C(\lambda_1) \cdots C(\lambda_M) \, . 
\end{equation}
We therefore define the impurity local magnetization 
$s_1^z$ by 
\begin{eqnarray} 
  {\frac 1 2} \langle \sigma_1^z \rangle_{g, h}  & = & {\frac 1 2} \langle \Psi_{g, h} | \,  \sigma_1^z |\Psi_{g, h} \rangle 
/ \langle \Psi_{g, h} | \Psi_{g, h} \rangle 
\nonumber \\  
& = & {\frac 1 2} 
{\frac  {\langle 0 |    
C(\lambda_1) \cdots C(\lambda_M) \, \sigma_1^{z}  \, B(\lambda_1) \cdots B(\lambda_M) | 0 \rangle} {\langle 0 |   
C(\lambda_1) \cdots C(\lambda_M) B(\lambda_1) \cdots B(\lambda_M) | 0 \rangle}} \, . 
\end{eqnarray} 
We now evaluate it through the expectation value of 
the local operator $e_1^{22}$ as follows. 
\begin{equation}  
{\frac 1 2 } \langle \sigma_1^z \rangle_{g, h} =  {\frac 1 2 } - \langle e_1^{22} \rangle_{g, h} 
\end{equation}
where $e_1^{a\, ,  b}$ denote the two-by-two matrices with only nonzero matrix element  
1 at  the entry of $(a, b)$ for $a, b=1, 2$. 
Hereafter, we call $\langle \sigma_1^z \rangle_{g, h}$ the local magnetization at the impurity site,  not $s_1^z= \langle \sigma_1^z \rangle_{g, h}/2$,  for simplicity. 

We introduce the Gaudin matrix $\Phi^{'}_{a, b}(\{ \lambda_{\alpha} \})$ with real parameters $z_k $ defined by $\xi_k=i \zeta/2 + z_k$ for $k=1, 2, \dots, N$, as follows.  
\begin{equation} 
\Phi^{'}_{a, b} = 2 \pi i N \left( \widehat{ \rho}(\lambda_a) \, \delta_{a, b}  
+{ \frac 1 N} a_2(\lambda_a - \lambda_b)  \right)  
\end{equation} 
where function  $\widehat{ \rho}(\lambda) $ is defined by  
\begin{equation} 
 \widehat{ \rho}(\lambda)  
= {\frac 1 N} \sum_{k=1}^{N} a_1(\lambda- z_k) - {\frac 1 N} \sum_{b=1}^{M} 
a_2(\lambda- \lambda_b)  \, .  
\end{equation} 
Putting the ground-state solution under a magnetic field, $\{ \lam_a \}$, with   
$z_1=x$ and $z_2=\cdots = z_N=0$, we have 
\begin{eqnarray} 
\widehat{ \rho}(\lambda)  & = & {\frac 1 N} \left( (N-1) a_1(\lambda) +  a_1(\lambda+x) \right)  - {\frac 1 N} \sum_{b=1}^{M} a_2(\lambda- \lambda_b)  \nonumber \\ 
 & = &  {\frac 1 N} \left( (N-1)  a_1(\lambda) +  a_1(\lambda+x) \right) 
- \int_{-B}^{B} a_2(\lambda- \lambda^{'})  d \lambda^{'}
+ O(1/N^2) \nonumber \\ 
& = & \rho(\lam) + O(1/N^2) \, .   
\end{eqnarray}
Here we recall that  $\rho(\lam)$ denotes the density of the Bethe roots for    
the ground-state solution under magnetic field $h$ satisfying (\ref{baeh}). 

Making use of the formula of the Quantum Inverse Scattering Problem \cite{Kitanine2000} 
\begin{equation}
e_1^{22} = D(\xi_1) (A(\xi_1)+D(\xi_1))^{-1}  
\end{equation}
and Slavnov's scalar product formula  we show 
\begin{eqnarray}  
\langle e_1^{22} \rangle_{g, h}
& = & \sum_{a=1}^{M} \det_M \left( (\Phi^{'} (\{ \lambda_{\beta} \} )^{-1}  
\Psi^{'}( \{ \lambda_{\beta} \} \setminus\{ \lambda_a \} \cup \{\xi_1 \})  \right) 
\nonumber \\  
& = & {\frac 1 N} \sum_{a=1}^{M} {\frac 1 {\rho(\lambda_a)}} \rho^{I}(\lambda_a, x) + O(1/N^2)  
\nonumber \\
& = & \int_{-B}^{B} \rho^{I} (\lambda, x) d \lambda + O(1/N^2) \, .    
\end{eqnarray}
Here $\rho^{I}(\lambda, x)$ denotes the solution of the integral equation 
\begin{eqnarray}\label{baeh-imp}
 \rho^{I}(z, x) + \int^B_{-B}a_2(z-z') \rho^{I}(z', x)dz'
 =a_1(z+x) \, ,  
\end{eqnarray}
and the matrix elements of the $M \times M$ matrix 
$\Psi^{'}(\{ \lambda_{\beta} \} \setminus\{ \lambda_a \} \cup \{\xi_1 \})$ are given by 
\begin{eqnarray}
\Psi^{'}_{b, c} & = & 
\left\{ 
\begin{array}{cc} 
2 \pi i \, a_1(\lambda_a +x ) & \mbox{if} \quad b = a \\ 
\Phi^{'}_{b, c } & \mbox{ otherwise } 
\end{array} 
\right. \, . 
\end{eqnarray}

%
%
\subsubsection{Analytic expression of the impurity local magnetization for small $|x|$}

Let us consider the case when the absolute value of the impurity parameter $x$  
is small enough with respect to a given value of magnetic field $h$. We assume that   
it is much smaller than the Fermi point $B$: $|x| \ll B$, where $B$ is related to 
the magnetic field $h$ by (\ref{eq:h-B}).  In this case we solve eq. \re{baeh-imp} by the Wiener-Hopf method. We define  $\rho^{I}_0(z, x)$ by  the solution of the integral equation  \re{baeh-imp} with $B=\infty$.  
Explicitly, we have $\rho_0^I(z, x) = \sigma_0(z+x)$.  
We define function $y^{I}(z,x)$ by 
\begin{equation} 
 y^{I}(z,x) =\rho^{I}(z+B, x) . 
\end{equation} 
Since the equation \re{baeh-imp} does not change by replacing $z$ and $x$ by $-z$ and $-x$, respectively,  we have $y^{I}(z, -x) =\rho^{I}(-z-B, x)$.  
We now define functions $\breve{y}_{\pm}^{I}(k, x)$ by 
\begin{equation} 
 \breve{y}_{\pm}^{I}(k,x) = \intall \mathcal{H}(\pm z)y^{I}(z,x)e^{ikz}dz . 
\end{equation}
By solving the integral equation 
\begin{eqnarray}\label{rho}
 \rho^I(z, x)=\rho_0^{I}(z, x)+\int_{|z'|>B} R(z-z')\rho^{I}(z', x)dz'  \label{eq:fund-I} 
\end{eqnarray}
we evaluate $\langle e_1^{22} \rangle_{g,h}$ through the following: 
\begin{eqnarray} 
 \int_{-B}^{B} \rho^{I} (z, x) d z 
& = & {\frac 1 2} -  \fr{\pi}{2\pi-2\z} \int_{|z|>B} \rho^{I}(z, x) dz \nonumber \\
 &=& {\frac 1 2}  - \fr{\pi}{2\pi-2\z} \lt \breve{y}_{+}^{I}(0,x)+\breve{y}_{+}^{I}(0,-x) \rt . 
\end{eqnarray} 
We thus obtain the analytic expression of the impurity local magnetization  
\begin{equation} 
\label{eq:s1-imp}
\langle \sigma_{1}^{z} \rangle_{g, h} 
= \fr{4 \z  }{\pi(\pi-\z) \sin \z} \cosh \left(\fr{\pi x}{\z} \right) \, h  + O(1/N^2) . 
\end{equation}
Here we recall that the magnetic field $h$ is related to the Fermi point $B$ by  (\ref{eq:h-B}).

The analytic expression \re{eq:s1-imp} of the impurity local magnetization 
 is consistent with the impurity susceptibility  (\ref{eq:sc-imp}). Therefore, it follows that    
the identification of the impurity contribution to the magnetic susceptibility in terms of the Bethe ansatz equations is valid. 
Here we remark that expression \re{eq:s1-imp} is valid only when the absolute value of impurity parameter $x$ is small enough such as $|x|< B$ so that we can apply the Wiener-Hopf method to solve the integral equation \re{baeh-imp}.  

%
\subsubsection{Pseudo-decoupled case}

Let us now consider the case when the absolute value of the impurity parameter $x$ is large. For instance, we assume that it is larger than the logarithm of the system size $N$: $|x| > \ln N$. It follows that  we have $\exp(-|x|) < 1/N$, and the energy gap due to the exchange interaction between the impurity site and the neighboring sites is smaller than the bulk excitation energy gaps,  which are of the order of $1/N$, 
and hence can be neglected with respect to them.

If  the absolute value of the impurity parameter $x$ is very large such as $x \gg \ln N$ 
and  $x \gg B$, we can show that the impurity local magnetization,  $\langle \sigma_{1}^{z} \rangle_{g, h}$, approaches the largest value 1  exponentially with respect to the impurity parameter $x$.  The right hand side of  the linear integral equation \re{baeh-imp}, which gives the source term to it,  is proportional to $\exp(-2x)$ if impurity parameter $x$ is very large such as $x \gg B$. Therefore, we have $\langle e_1^{22} \rangle \sim \exp(-2x)$ for large values of $x$,  which leads to the following:    
\begin{equation} 
1 - \langle \sigma_1^{z} \rangle_{g,h} \sim \exp(-2x) \quad \mbox{for} \, \,  x \gg 
B \, \, \mbox{and} \, \,  x \gg \ln N  \, . \label{eq:exp}
\end{equation}

We now show that the impurity local magnetization becomes easily saturated even with an infinitesimally small magnetic field, if pseudo-decoupling occurs. In fact, even if the absolute value of the impurity parameter $x$ is very large such as $x \gg \ln N$, we can take a very small magnetic field $h$ such that the value of the Fermi point $B$ is much larger than $|x|$. Then, we can apply the Wiener-Hopf method to the integral equation \re{baeh-imp}, and we obtain the expression \re{eq:s1-imp} of the impurity local magnetization for a very small magnetic field $h$ with a large value of $x$.  

Let us estimate the small magnetic field $h$ for which the expression \re{eq:s1-imp} is valid in the case of pseudo-decoupling.  If we determine the magnetic field by $2h(M) = E(M-1) - E(M)$ in the $N$-site chain with $M-1$ down spins,  then $h(M)$ is given by of the order of $1/N$. However, the magnetic field is much smaller than $O(1/N)$ if 
pseudo-decoupling occurs. Let us now assume the following conditions:  
\begin{equation}
 1/N  >  \exp(-x) >  \exp(-B)  \, ,  \quad  (h)^{1/p_0}  \propto \exp(-B) \, , 
\end{equation}   
where $p_0=\pi/\zeta$. For instance,  $p_0=3$ for $\zeta=\pi/3$.  We suggest that  
the magnetic field $h$ which triggers the impurity local magnetization is given by $h < 1/N^3$, which is very small.  We thus expect that if we add a small magnetic field $h$ such as $h \sim 1/N^3$, then the local magnetization at the impurity site readily approaches the largest value 1. 

The result is consistent with another viewpoint that the impurity spin becomes a free spin if the exchange coupling with other spins vanishes. It is clear that a free spin becomes saturated with an infinitesimally small magnetic field.  

%
\subsubsection{Numerical estimates of the impurity local magnetization}

\begin{figure}[h]
 \centering
 \includegraphics[width=8cm]{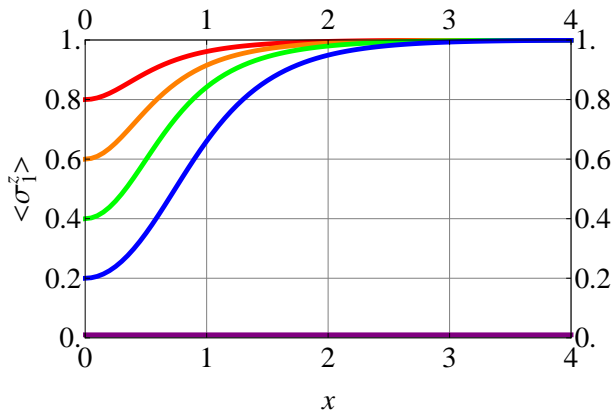}
 \caption{$\vtop{
\hbox{ Local magnetization at the impurity site, $\langle \sigma_1^{z} \rangle_{g,h}$, at $T=0$ and $\zeta=\pi/3$ }
\hbox{ versus impurity parameter $x$ for different values of magnetic field (purple,  }
\hbox{ blue, green, orange, and red, in increasing order) for $N=10$. }
\hbox{ The purple line corresponds to the case of the zero magnetic field. }
}$}
\label{fig-imp}
\end{figure}

Let us now confirm the analytic results of the impurity local magnetization by numerical data. 
We plot the numerical estimates of the local magnetization at the impurity site,  
$\langle \sigma_1^{z} \rangle_{g,h}$, against the impurity parameter $x$ in the case of $N=10$ in Fig. \ref{fig-imp}. The estimates are evaluated by numerically diagonalizing the Hamiltonian  $\xxz(x)$ with $N=10$. 

For small values of the impurity parameter $x$, the impurity local magnetization, $\langle \sigma_1^{z} \rangle_{g, h}$, is proportional to the magnetic field $h$, and it increases with respect to $x$ similarly as $\cosh \pi x/\zeta$.  However,  for large values of the impurity parameter $x$, it approaches the largest value 1.0.  Moreover, under an infinitesimally small magnetic field $h$, the impurity spin 
takes very quickly a positive large value less than 1.0 .  We also observe that it vanishes under zero magnetic field for any values of the impurity parameter $x$.

%
%
\subsection{Numerical estimates of the impurity susceptibility}

We now evaluate  numerically the magnetic susceptibility by calculating the energy eigenvalues through solutions of the Bethe ansatz equations. 
For a given number $M$ satisfying $M \le N/2$  we solve the BAE \re{bae} for the ground state with $M$ down spins,  and evaluate the ground state energy $E(M)$ through the solution by making use of eq. \re{en}. 
We also evaluate the energy $E(M-1)$ for the ground state with $M-1$ down spins. 
Then, we determine the magnetic field $h(M)$ by the following relation: 
\begin{equation} 
2 h(M) = E(M-1)- E(M) \, . \label{difference} 
\end{equation} 
Here we recall that the total magnetization $S^z_{tot}$ is given  by $S^z_{tot}= (N- 2M)/2$ in the case of  $M$ down spins.  We define the total susceptibility $\chi_{tot}(M)$ under magnetic field $h(M)$ by 
\begin{equation}
\chi_{tot}(M)  =  (2S^z_{tot}(M-1)-2S^z_{tot}(M))/(h(M-1)-h(M)) \, . \label{eq:chiM}
\end{equation}  
We can evaluate the susceptibility at $h=0$, i.e. $\chi_{tot}(M)$ for $M=N/2$, 
by taking the difference through eq. (\ref{eq:chiM}). 
It is close to the definition \re{eq:def-chi} of the susceptibility by the derivative of 
$2S^z_{tot}$ with respect to magnetic field $h$.    
The susceptibility per site $\chi(x)$ is given by $\chi(x)=\chi_{tot}/N$.  We then 
evaluate the impurity susceptibility $\chi_{imp}(x)$ 
from the total susceptibility  $\chi_{tot}$ by the following relation: 
$\chi_{tot} = (N-1) \chi(0) + \chi_{imp}(x)$.  
\footnote{Numerically we have evaluated  the susceptibility $\chi_{tot}$ at $h=0$ 
and then the impurity susceptibility $\chi_{imp}(x)$ at $h=0$ as follows. We make a graph of magnetization $2S^Z=N-2M$ versus $h(M)$, and interpolate the data points of $2S^z$ versus $h(M)$ as a polynomial of $h$ of some degree.  We employ the ratio  $2S^z_{tot}(M-1)/h(M-1)$ at $M=N/2$ as  the susceptibility  $\chi_{tot}$  at $h=0$ assuming that $S^z_{tot}(M)= 0$ and $h(M)=0$ for $M=N/2$. 
Here we remark that the total susceptibility  $\chi_{tot}$ is $O(N)$, i.e. of the order of the system size $N$. We repeat the evaluation of  $\chi_{tot}$ for different values of the system size $N$, and plot the total susceptibility divided by $N$,  $\chi_{tot}/N$, against the inverse of the system size, $1/N$.  Then,  we interpolate the data points of  $\chi_{tot}/N$ versus $1/N$ as a polynomial of $y=1/N$ of some degree and employ the coefficient of the $y^1$ term as the shift $\D \chi= \chi_{imp}(x) - \chi(0)$. By adding the susceptibility per site with no impurity, $\chi(0)$, to it,  we obtain  $\chi_{imp}(x)$.  
}

\begin{figure}[h]
\begin{minipage}{0.50\hsize}
  \begin{center}
   \includegraphics[width=6cm]{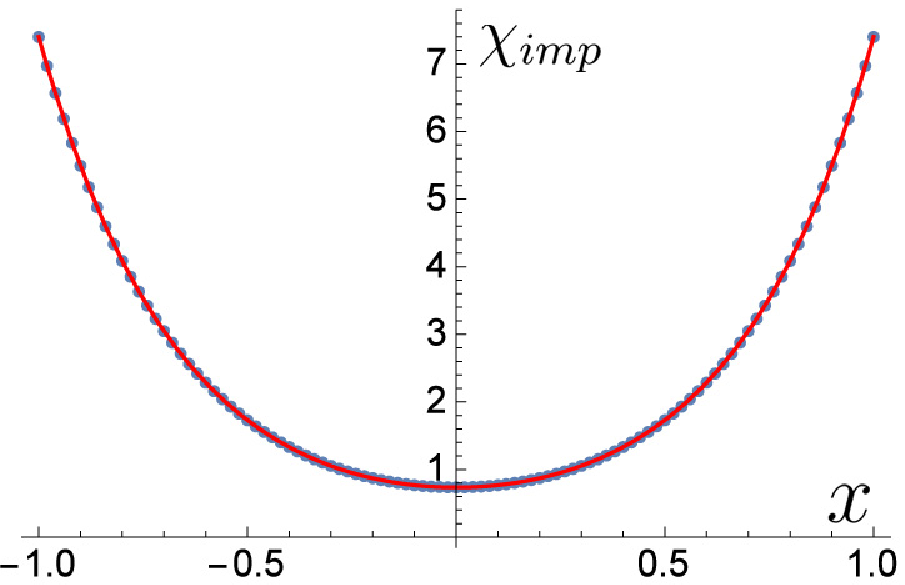} 
  \end{center}
  \caption{$\vtop { 
\hbox{Impurity susceptibility  $\chi_{imp}(x)$} 
\hbox{ versus impurity $x$ with small $x$ }
\hbox{for $\z=\fr{\pi}{3}$.  Extracted from} 
\hbox{9 data points from $N=30$}
\hbox{to $100$ by 10 and $N=\infty$. }
\hbox{Red curve: analytic formula \re{eq:sc-imp} } 
\hbox{Blue dots: numerical estimates.  }
}$ } 
  \label{fchi2}
 \end{minipage}
 \begin{minipage}{0.70\hsize}
  \begin{center}
   \includegraphics[width=6cm]{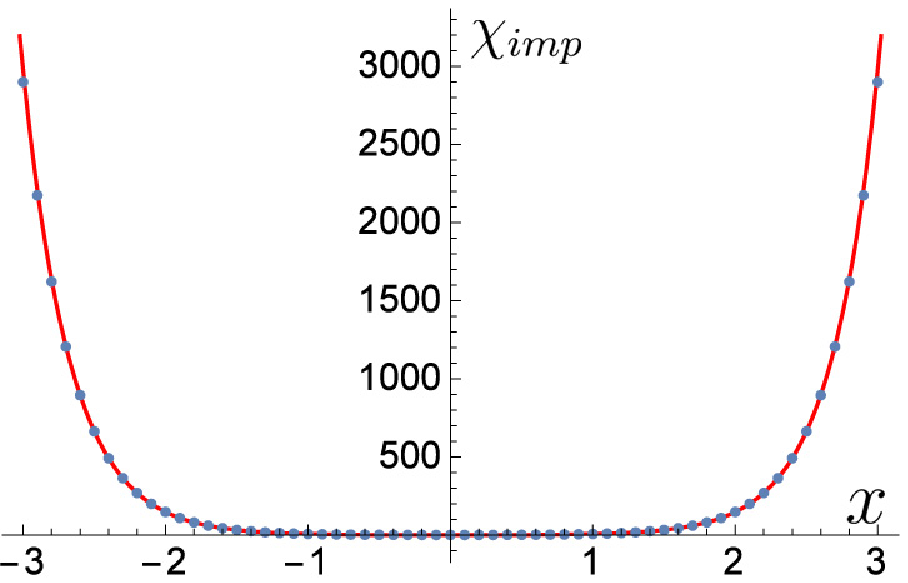} 
  \end{center}
  \caption{$\vtop{
\hbox{Impurity susceptibility $\chi_{imp}(x)$ } 
\hbox{versus $x$ in a wide range of $x$ }
\hbox{for  $\z=\fr{\pi}{3}$. Extracted from} 
\hbox{5 data points from $N=2,000$}
\hbox{to $10, 000$ by 2, 000 and  $N=\infty$. } 
\hbox{Red curve: analytic formula \re{eq:sc-imp} } 
\hbox{Blue dots: numerical estimates.  }
}$ }
  \label{fchi1}
 \end{minipage}
\end{figure}

The numerical estimates of the impurity susceptibility $\chi_{imp}(x)$ at $h=0$ are plotted against impurity parameter $x$ in Figs. \ref{fchi2} and \ref{fchi1}.
The impurity susceptibility increases with respect to the absolute value of impurity parameter $x$, which is due to the Kondo effect. 
The numerical estimates  (blue dots) are consistent with the analytic formula \re{eq:sc-imp} (red line) in Figs. \ref{fchi2} and \ref{fchi1}. Here,  the typical system size $N$ for the data of Figs. \ref{fchi2} and \ref{fchi1} is given by $N=10^2$ and $10^4$, respectively.  
We suggest that if the system size $N$ is large enough, then the numerical estimates of the impurity susceptibility $\chi_{imp}(x)$ at $h=0$
are consistent with the analytic formula  \re{eq:sc-imp}.   
For instance,  the numerical estimates of $\chi_{imp}(x)$ for $N=100$ 
shown in Fig. \ref{fchi2} only from $x= -1$ to 1 are smaller than the corresponding values  of the analytic formula  \re{eq:sc-imp} if  the absolute value of impurity parameter $x$ is larger than 2: $|x| > 2$, while those of  $N=10^4$ are rather close to  the analytic curve  \re{eq:sc-imp} even for $|x|=3$ as shown in Fig. \ref{fchi1}.

By applying the Wiener-Hopf method formulated in Ref. \cite{YY} we can show that the $x$-dependence of the susceptibility per site $\chi(x)$ changes if we add higher-order terms with respect to the expansion parameter $\exp(-\pi B/ \zeta)$. 
However, it seems that the same $x$-dependence as in eq. \re{eq:sc-imp} is reproduced 
for the impurity susceptibility at zero magnetic field since higher-order terms as in Ref. \cite{YY} do not remain when $h$ is small.

In Ref.\cite{Sch}, it is stated that the impurity susceptibility $\chi_{imp}$ should depend on the parameter $x$ such as $\exp\left(\pi |x|/{\z} \right)$  in terms of the notation of the present paper, 
which is different from $\cosh\left( \pi x/{\z} \right)$. However, the exponential dependence is not exact as an analytic solution in the Wiener-Hopf method.  
Furthermore,  it is not consistent with the numerical estimates of   the impurity susceptibility $\chi_{imp}$ even for small $x$ (see also Fig. \ref{fchi2}).

%
%

%
%
\setcounter{equation}{0}
\section{ Specific heat due to impurity}\label{spe}

\subsection{Thermal Bethe ansatz}

We now evaluate the specific heat by solving the truncated integral equations 
of  the thermal Bethe ansatz \cite{TS,T73,FZ,T}. 
We show explicitly how  the specific heat depends on the impurity parameter $x$.  
We restrict the anisotropy parameter $\z$ as $\z=\fr{\pi}{3}$ throughout this section.

Let us consider the string solutions for the BAE at $\z=\fr{\pi}{3}$. 
It is known that there are only three different types of solutions $z_j \ (j=1,2,3)$ 
for $\z=\fr{\pi}{3}$. Explicitly, we have 
\begin{eqnarray}
 \left\{
 \begin{array}{lll}
  z_1=z_1^R \\
  z_{2, \, \pm}=
   \begin{cases}
    z_2^R + \fr{i\z}{2} \\
    z_2^R - \fr{i\z}{2} 
   \end{cases} \\
  z_3=z_3^R+\fr{\z}{2}p_0 i, \ \ \ \ \ \ p_0=\fr{\pi}{\z},
 \end{array}
 \right.
\end{eqnarray}
where each $z_j^R$ gives the real part or the center of the string solution $z_j$. 
We remark that $\sh(z \pm \fr{\z}{2}p_0 i)= \pm i \ch z$. 
In terms of string solutions the BAE is now  expressed as follows. 
\begin{eqnarray}\label{tbae}
 \fr{1}{N} \{ a_j(z+x)+(N-1)a_j(z) \} & = & \sign (q_j) (\rho_j(z)+\rho_j^h(z))
 +\sm{k=1}{3} T_{jk}*\rho_k(z) \nonumber \\ 
& & {\mbox{for}} \, \, j = 1, 2, 3.    
\end{eqnarray}
Here, unknown functions $\rho_j(z)$  and $\rho_j^h(z)$ denote 
the particle and hole densities, respectively, with respect to the real part $z_j^R$ of the string solutions $z_j$, and the symbol $f*g$  denotes 
the convolution of given functions $f$ and $g$. 
The functions $a_j(z)$ for $j=1,2, 3$ are given by 
\begin{equation} 
 a_j(z)=\fr{1}{2\pi}\fr{2\sin \z q_j}{\ch 2z + \cos \z q_j} \, , 
\end{equation} 
where $q_j$ are given by 
\begin{equation} 
q_1=2, \ \ q_2=1, \ \ q_3=-1 \, . 
\end{equation} 
The matrix $T$ is defined by  
\begin{equation}
 T_{jk}=
 \begin{pmatrix}
  a_2 & a_1 & -a_1 \\
  a_1 & a_2 & -a_2 \\
  -a_1 & -a_2 & a_2
 \end{pmatrix}.
\end{equation}
Here we remark that $a_1$ and $a_2$ are the same functions,  
as defined by \re{eq:an}.  
By making use of the identity
\begin{eqnarray}
 a_2=-a_3
\end{eqnarray}
we reduce the number  of equations of \re{tbae} from 3 to 2.

Let us now express physical quantities in terms of $\rho_j(z)$ and 
$\rho_j^h(z)$. 
The energy of this system is described by
\begin{equation}
 e(x, T) = \k \sm{j=1}{3} \intall a_j(z)\rho_j(z) dz \, , 
\end{equation}
where $\k$ is given by 
\begin{equation}
 \k = -\pi \sin \z,
\end{equation} 
and the entropy is expressed as 
\begin{eqnarray}
 s=\sm{j=1}{3} \intall\lt \rho_j(z)\log \[1+\fr{\rho_j^h(z)}{\rho_j(z)}\]
 +\rho_j^h(z)\log \[1+\fr{\rho_j(z)}{\rho_j^h(z)}\] \rt dz.
\end{eqnarray}
Let us take  the free energy $f=e-Ts$ ($T$ : temperature) with respect to $\rho_j$ 
under the conditions \re{tbae}. Then, it is clear that the thermal equilibrium condition $\delta f=0$ is equivalent to
\begin{equation}
 \log \eta_j = \fr{\k}{T}a_j
 + \sm{k=1}{3}\sign(q_k)T_{jk}*\log(1+\eta_j^{-1}), \label{tec}  
\end{equation} 
where functions $\eta_j$ are defined by 
\begin{equation} 
 \eta_j = \fr{\rho_j^h}{\rho_j}.
\end{equation}
The impurity parameter $x$ does not remain 
after by taking the variation with respect to $\rho_j$ and $\rho_j^h$ under the constraint of BAE \re{tbae}, 
and hence the above equations are the same as those of the homogeneous 
spin-1/2 XXZ spin chain. 
By putting them back to the free energy, its final form is given by the following: 
\begin{eqnarray}
 f=-\fr{T}{N} \sm{j=1}{3} \sign (q_j) \intall
 \lt (N-1)a_j(z)+a_j(z+x) \rt \log (1+\eta_j(z)^{-1}) dz.
\end{eqnarray}
Equations \re{tec} are solved numerically \cite{FZ}. As we said before, $j=2$ and $j=3$ indicate the same equation. 
Only two of the three eta-functions are independent and the equations can be reformulated as follows. 
\begin{eqnarray}\label{heq}
 \left\{
 \begin{array}{lll}
 h_1=\log \[ 1+\exp \[-\fr{\k}{T}b_1-2b_2*h_1-2b_1*h_2  \] \] \\
 h_2=\log \[ 1+\exp \[-\fr{\k}{T}b_2-b_1*h_1-2b_2*h_2  \] \]
 \end{array}
 \right.,
\end{eqnarray}
where $h_j $ and the Fourier transforms of $b_j$ are given by  
\begin{eqnarray}
 h_j(z) &=& \log (1+\eta_j^{-1}(z)), \\
 \til{b}_1(\o) &=& \fr{\til{a}_1}{1-\til{a}_2},
 \ \ \ \ \ \ 
 \til{b}_2(\o) = \fr{\til{a}_2}{1-\til{a}_2} . 
\end{eqnarray}
Explicitly we have 
\begin{eqnarray}
 b_1(z) &=& \fr{3 \ch \fr{3z}{2}}{\sqrt{2} \pi \ch 3z},
 \ \ \ \ \ \ 
 b_2(z)=\fr{3}{4\pi \ch \fr{3z}{2}}.
\end{eqnarray}
Here, the expressions of functions $b_j$ are specific to $\z=\fr{\pi}{3}$. In terms of these functions, the free energy is written by
\begin{eqnarray}
 &&f = -\fr{T}{N} \intall \lt \a_1 h_1
 +\a_2 \(2h_2+ \fr{\k}{T}b_2+b_1*h_1+2b_2*h_2 \) \rt dz, \\
 &&\a_j(z) = (N-1)a_j(z)+a_j(z+x).
\end{eqnarray}
The specific heat is calculated by the same approach. Let us set the following quantities:  
\begin{eqnarray}
 u_j=T^2 \pbib{h_j}{T}, \ \ \ \ \ v_j:=\pbib{u_j}{T}.
\end{eqnarray}
Then, the specific heat ($c(x, T)=\pbib{e}{T}$, $e=-T^2\pbib{}{T}\(\fr{f}{T}\)$) becomes
\begin{eqnarray}
 c(x, T)=\fr{1}{N} \intall \lt \a_1 v_1
 +\a_2 \(2v_2+b_1*v_1+2b_2*v_2 \) \rt dz.
\end{eqnarray}
During the procedure, we also have its energy
\begin{eqnarray}
 e=\fr{1}{N} \intall \lt \a_1 u_1
 +\a_2 \(2u_2-\k b_2+b_1*u_1+2b_2*u_2 \) \rt dz.
\end{eqnarray}
The equations for $u_j$ are given by 
\begin{eqnarray}
 \left\{
 \begin{array}{lll}
 u_1=\(1-e^{-h_1}\) \[-\k b_1-2b_2*u_1-2b_1*u_2  \] \\
 u_2=\(1-e^{-h_2}\) \[-\k b_2-b_1*u_1-2b_2*u_2  \]
 \end{array}
 \right., 
\end{eqnarray}
and  those for $v_j$  by 
\begin{eqnarray}
 \left\{
 \begin{array}{lll}
 v_1=\fr{1}{T^2}\fr{{u_1}^2}{e^{h_1}-1}
 +\(1-e^{-h_1}\) \[-2b_2*v_1-2b_1*v_2  \] \\
 v_2=\fr{1}{T^2}\fr{{u_2}^2}{e^{h_2}-1}
 +\(1-e^{-h_2}\) \[-b_1*v_1-2b_2*v_2   \].
 \end{array}
 \right.
\end{eqnarray}

%
%
\subsection{Specific heat with an impurity near zero temperature}

Let us consider  the impurity specific heat, $c_{imp}(x,T)$,  i.e., the specific heat on the impurity site with parameter $x$ at temperature $T$.  We define it by  
\begin{equation} 
N c(x, T)=(N-1)c(0, T)+c_{imp}(x, T) \, .     
\end{equation}
Here  the specific heat per site is denoted by $c(x, T)$. 
We remark that the specific heat of the total system $C_{tot}(x, T)$ is given by 
$N c(x, T)$: $C_{tot}(x, T)=N c(x, T)$. 

If  temperature $T$ is very low and the absolute value of impurity parameter  $x$ 
is small enough with respect to the temperature,
we suggest the analytical expression of $c_{imp}(x,T)$ from   
that of the impurity magnetic susceptibility  $\chi_{imp}(x, T)$ as     
\begin{equation}\label{eq:cimp-lowT}
 c_{imp}(x,T)   = \fr{2 \z T}{3\sin \z} \ch \fr{\pi x}{\z} .
\end{equation}

\begin{figure}[ht]
\begin{minipage}{0.5\hsize}
  \begin{center}
   \includegraphics[width=6cm]{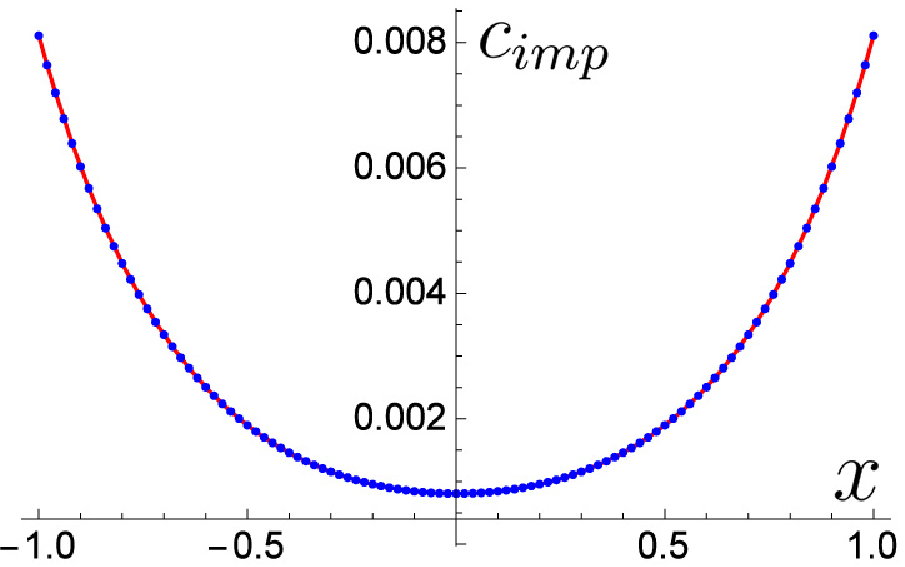}
  \end{center}
  \caption{$\vtop {
  \hbox{Impurity specific heat $c_{imp}(x, T)$ }
  \hbox{ versus $x$ in a narrow range}  
  \hbox{ of $x=0$ at  $T=10^{-3}$ for $\z=\fr{\pi}{3}$.}
  \hbox{Red curve is given by eq. \re{eq:cimp-lowT}. }
  }$}
  \label{fcfit}
 \end{minipage}
\begin{minipage}{0.5\hsize}
  \begin{center}
   \includegraphics[width=6cm]{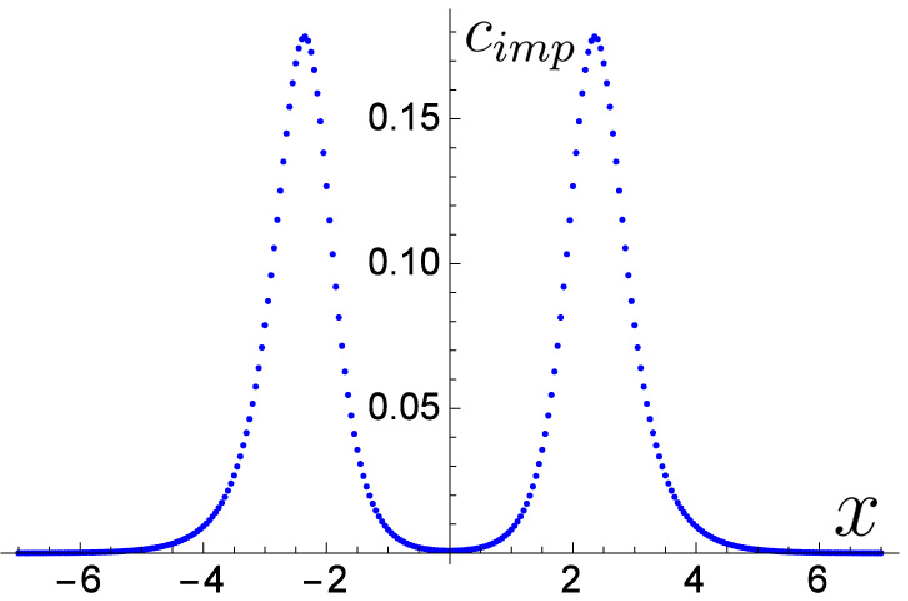}
  \end{center}
  \caption{$\vtop { 
\hbox{Impurity specific heat $c_{imp}(x, T)$}
\hbox{ versus impurity $x$ at  $T=10^{-3}$ } 
\hbox{for $\z=\fr{\pi}{3}$.} } $}
  \label{fc1}
 \end{minipage}
\end{figure}

In Fig. \ref{fcfit} the impurity specific heat $c_{imp}(x, T)$ at low temperature $T=0.001$ 
is plotted against impurity parameter $x$. 
Here, the numerical estimates of the impurity specific heat $c_{imp}(x, T)$  in the interval from $x=-1.0$ to  $1.0$  near the origin of $x=0$   
are fitted completely well as a function of $x$  to the analytic expression \re{eq:cimp-lowT}.  
There is no fitting parameter. 
The impurity specific heat \re{eq:cimp-lowT} reduces to the low-temperature result of the homogeneous XXZ spin chain (see e.g. \cite{T73}),  when we put $x=0$.
We suggest that the expression \re{eq:cimp-lowT} of the impurity specific heat 
at low $T$ should be valid for any value of $\z$ in the massless regime, 
although we evaluated it only for $\z=\fr{\pi}{3}$.

The impurity specific heat $c_{imp}(x,T)$ is plotted against parameter $x$ in a wider range of $x$ in Fig. \ref{fc1}. At $x=0$ the graph is convex and almost flat.  
It increases with respect to $|x|$, and reaches two peaks at $x= \pm 2.36$ .  
We observe in Fig. \ref{fc1} that the impurity spin becomes close to a free spin at $T=10^{-3}$  for large values of  $|x|$ such as $|x| > 2.36$:    
The impurity specific heat $c_{imp}(x, T)$ approaches zero as $|x|$ increases. 
Here we remark that the specific heat of a free spin vanishes 
under zero magnetic field at any temperature.

We suggest that the peak positions $x = \pm x_p$ in the graph of  
$c_{imp}(x, T)$ against parameter $x$ is approximately given by 
the solution of the following equation  
\begin{equation} 
T = 1/\cosh {\frac {\pi x} {\zeta}}  \, . \label{eq:crossoverT}
\end{equation} 
For $T=10^{-3}$ we have $x=2.53$ with $\zeta=\pi/3$.   
For $T=10^{-2}$, $c_{imp}(x, T)$ has two peaks with $x_p=1.59$, 
while the solution of eq. \re{eq:crossoverT}  is given by $x=1.77$.

%
%
\subsection{Impurity specific heat at finite temperatures}

\begin{figure}[ht]
\begin{minipage}{0.5\hsize}
  \begin{center}
   \includegraphics[width=6cm]{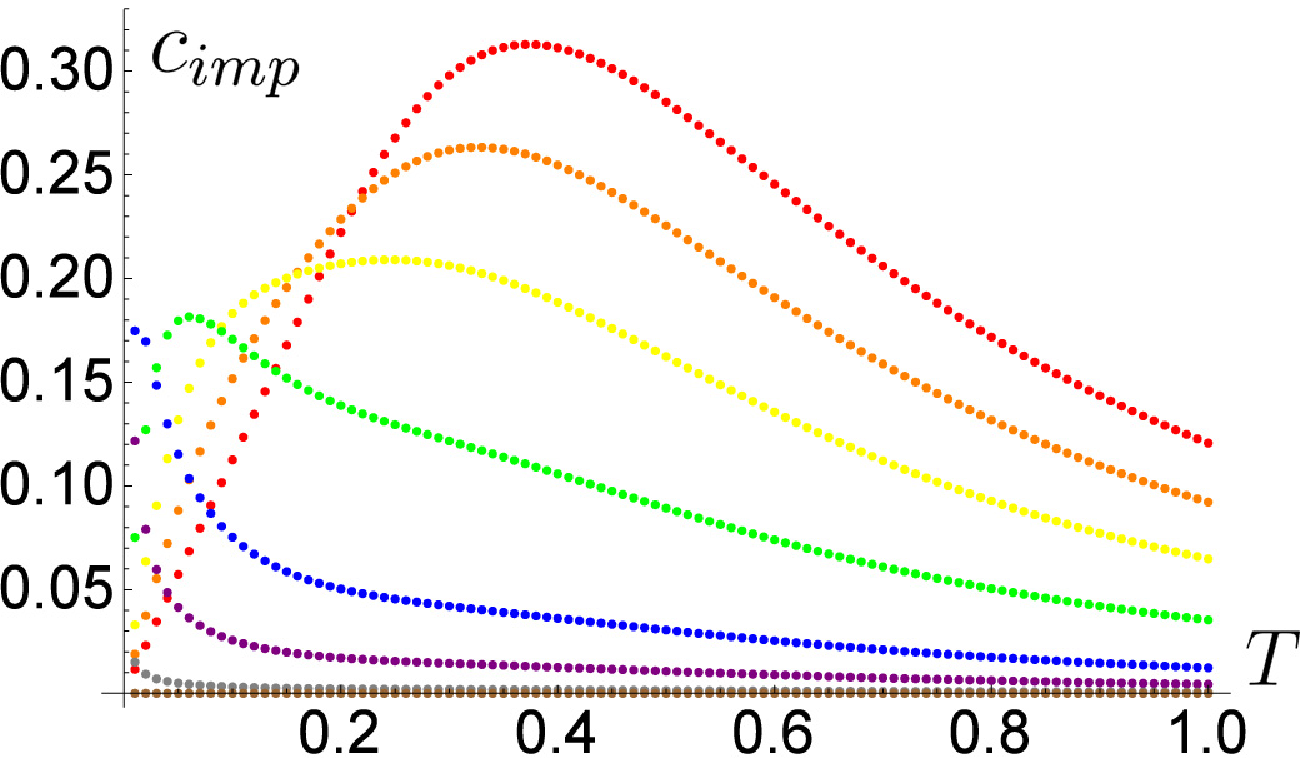}
  \end{center}
  \caption{$\vtop {
\hbox{Impurity specific heat $c_{imp}(x, T)$ }
 \hbox{versus temperature $T$ for $\z=\fr{\pi}{3}$,}
  \hbox{and $x=$0.3 (red), 0.5 (orange),}
  \hbox{0.7 (yellow), 1.0 (green),1.5 (blue),}
  \hbox{2.0 (purple), 3.0 (gray), 10.0 (brown).}
  }$}
  \label{fcimp}
 \end{minipage}
 \begin{minipage}{0.5\hsize}
  \begin{center}
   \includegraphics[width=6cm]{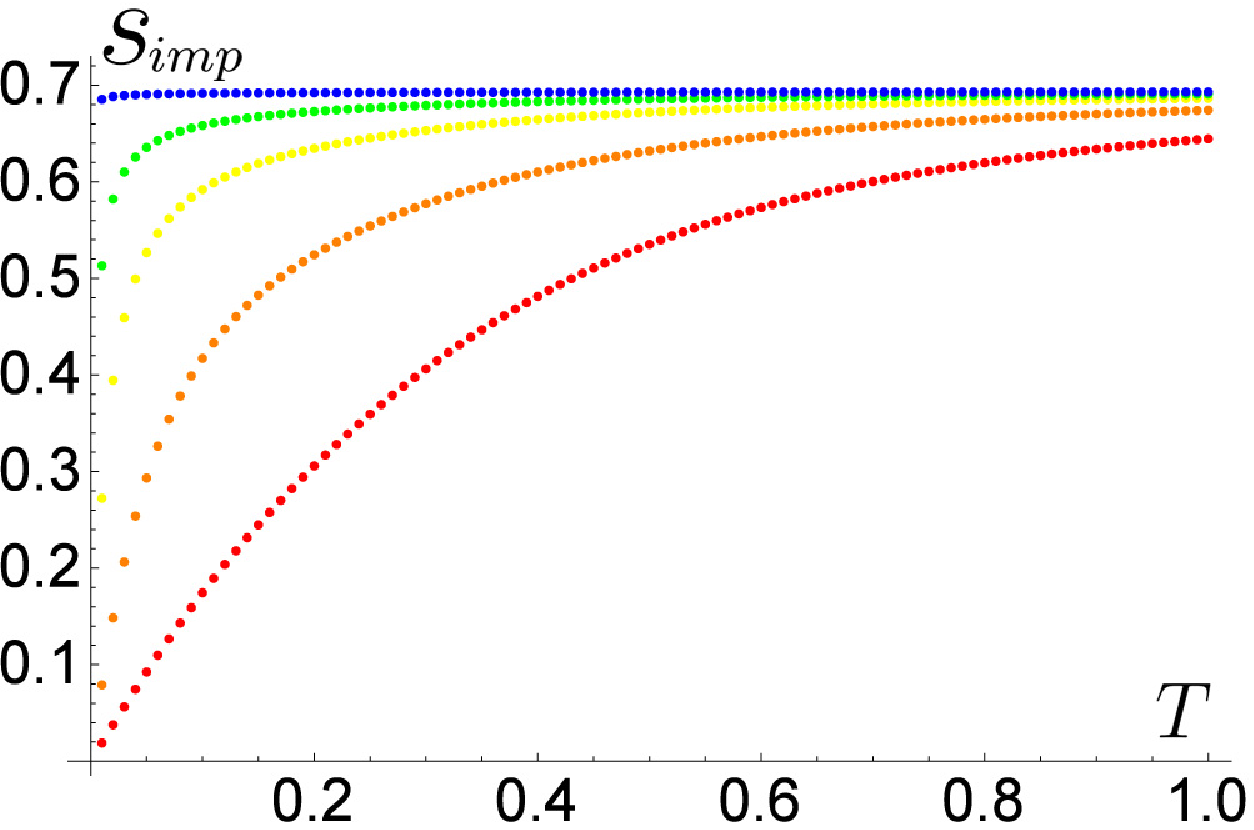}
  \end{center}
  \caption{$\vtop {
\hbox{Impurity entropy $s_{imp}(x, T)$ }
\hbox{ versus temperature $T$ }
  \hbox{$\z=\fr{\pi}{3}$, $x=$0.5 (red),}
  \hbox{1.0 (orange), 1.5 (yellow),}
  \hbox{2.0 (green), 3.5 (blue).}
  }$}
  \label{fsimp}
 \end{minipage}
\end{figure}

We shall show that the impurity spin has two aspects in finite temperatures: 
the Kondo effect in low temperature  and the tendency to become a free spin  through pseudo-decoupling in high temperature. 
Similarly as the impurity specific heat $c_{imp}(x,T)$, we define 
the impurity entropy $s_{imp}(x, T)$ by 
\begin{equation} 
N s(x, T)=(N-1) s(0, T)+s_{imp}(x, T) \, .  
\end{equation}
Here, the entropy of the total system,  $S_{tot}(x, T)$, is given by $S_{tot}(x, T)=N s(x, T)$. We also remark that we evaluate the entropy by $s=(e-f)/T$ from the energy $e$ and the free energy $f$.

The numerical estimates of the impurity specific heat $c_{imp}(x, T)$ 
and the impurity entropy $s_{imp}(x, T)$ are plotted in 
Figs. \ref{fcimp} and \ref{fsimp}, respectively, against temperature $T$.  
We observe the following: (i) The impurity specific heat $c_{imp}(x,T)$ has a peak approximately at $T=0.4$ in Fig.\ref{fcimp} for small absolute values of  parameter $x$ such as $|x|=0.3$;  
(ii)  The impurity specific heat $c_{imp}(x, T)$ grows linearly with respect to the temperature $T$ near zero temperature, and the gradient of the linear specific heat becomes larger 
as the impurity parameter $x$ increases; (iii) The peak position of the impurity specific heat  $c_{imp}(x, T)$  
shifts to lower temperatures as the impurity parameter $x$ increases;  (iv) The peak value of the impurity specific heat  $c_{imp}(x,T)$ decreases as the peak position shifts to lower temperatures.

We suggest that the Kondo effect leads to the peaks in the graphs of the impurity specific heat $c_{imp}(x,T)$ versus temperature $T$ in Fig. \ref{fcimp}. We recall that the Bethe ansatz equations \re{bae} in the XXX limit with $x=1$ 
are similar to those of the Kondo model \cite{Kondo}. 
Furthermore,  it seems that the increase of the gradient of the impurity specific heat $c_{imp}(x, T)$ at low temperature is consistent with the analytic expression \re{eq:cimp-lowT}.

In high temperatures such as $T =1.0$,  the impurity entropy $s_{imp}(x,T)$ approaches the value $\ln 2 \approx 0.69$ for any value of $x$, as shown in Fig. \ref{fsimp}. It is standard that the impurity spin becomes independent of the other spins in the chain at  high temperature.

%
%
\subsection{Crossover temperature}

Let us now determine the crossover temperature between the two regimes: the low-temperature regime where the Kondo effect is dominant and the high-temperature regime where the impurity spin tends to become a free spin through pseudo-decoupling. 

\begin{figure}[ht]
  \begin{center}
   \includegraphics[width=8cm]{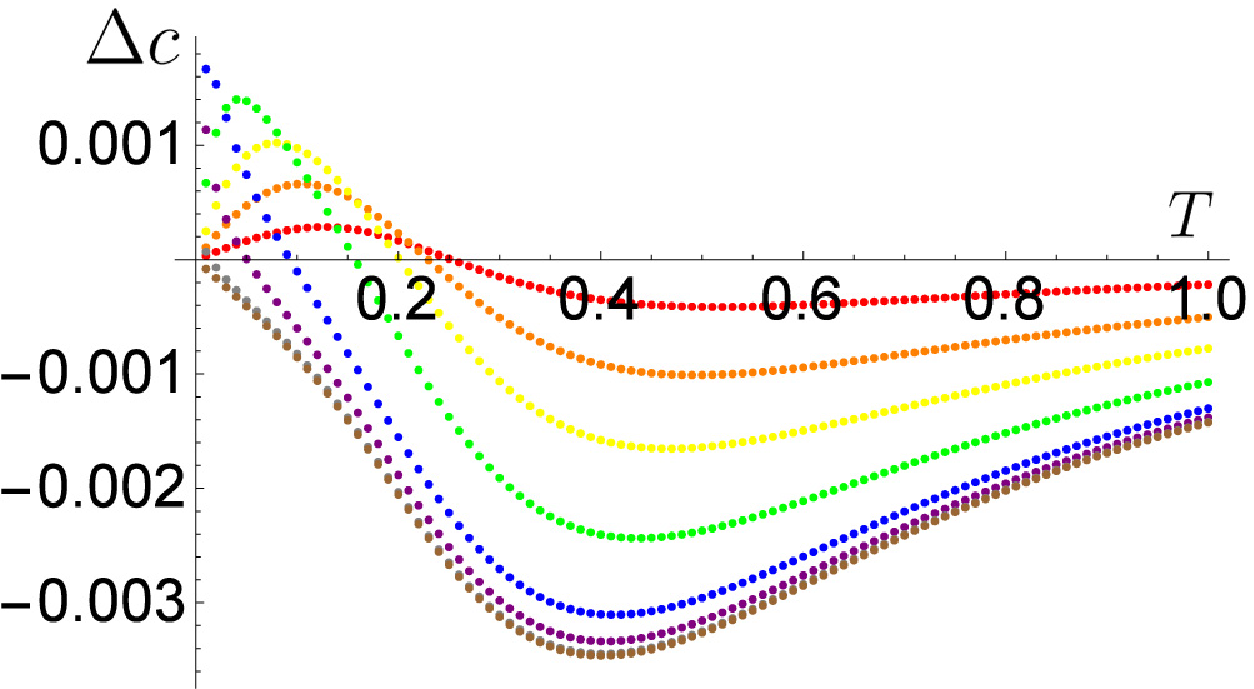}
  \end{center}
  \caption{$\vtop {
  \hbox{Specific heat shift $\D c(x, T)$}
  \hbox{versus temperature $T$ }
  \hbox{for $\z=\fr{\pi}{3}$, $N=100$: }
  \hbox{$x=$0.3 (red), 0.5 (orange), 0.7 (yellow),}
  \hbox{1.0 (green),1.5 (blue), 2.0 (purple),}
  \hbox{3.0 (gray), 10.0 (brown).}
  }$}
  \label{ftc}
\end{figure}

By the difference: $\D c(x,T) = c(x,T)-c(0,T)$, 
we define the shift of the specific heat per site with impurity parameter $x$ from that of the homogeneous chain  at temperature $T$.  
Here we have   from eq. \re{eq:cimp-lowT}  
the following expression in low temperature: 
\begin{equation}\label{cshift}
\D c(x,T)  =  \fr{2 \z T}{3N\sin \z} \( \ch \fr{\pi x}{\z}-1\) . 
\end{equation}
The numerical estimates of  the specific heat shift  $\D c(x, T)$ are plotted 
against temperature $T$ in Fig. \ref{ftc}.  Each dotted curve shows the graph of the data
of $\D c(x, T)$ with a fixed value of impurity parameter $x$.  

\begin{figure}[h]
\begin{minipage}{0.55\hsize}
  \begin{center}
   \includegraphics[width=7cm]{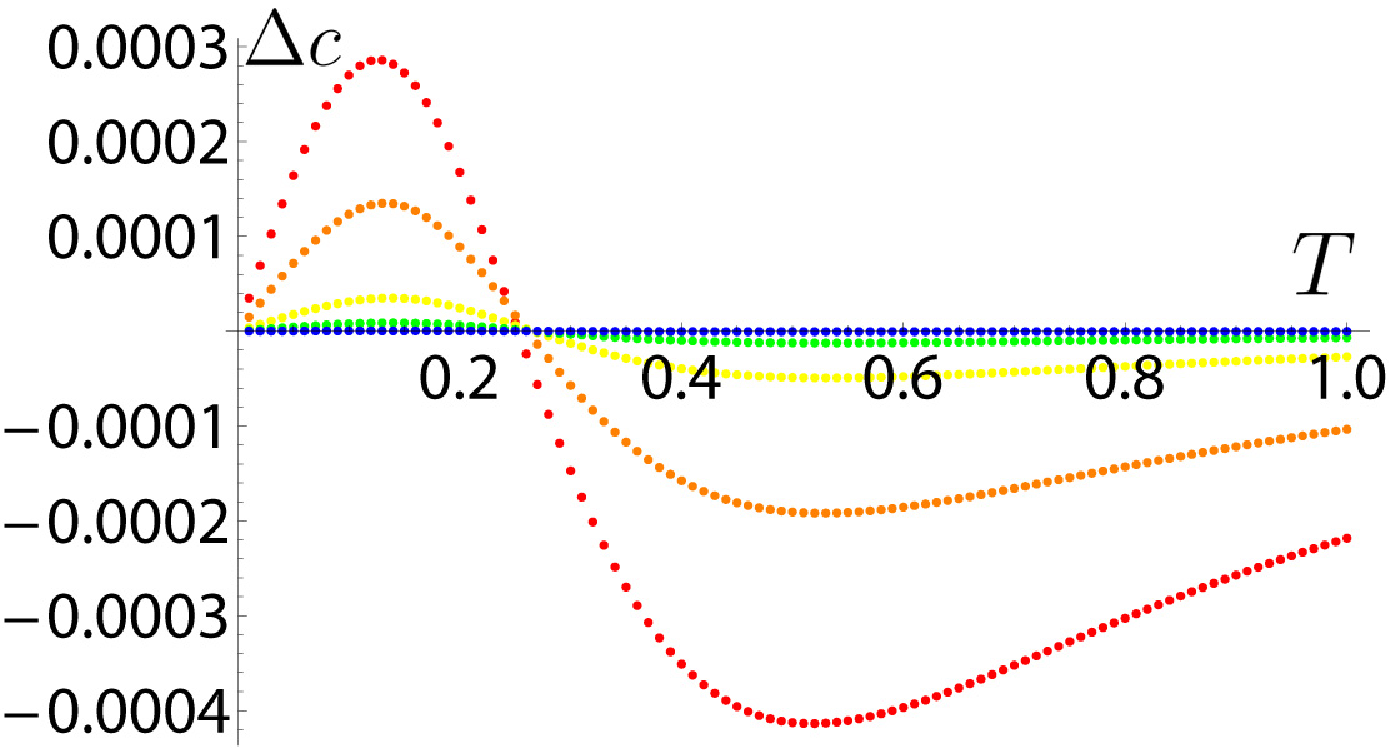}
  \end{center}
  \caption{$\vtop {
   \hbox{Specific heat shift $\D c$ with }
  \hbox{small values of impurity $x$}
  \hbox{for $\z=\fr{\pi}{3}$ and $N=100$.}
  \hbox{$x=$0.3 (red), 0.2 (orange),}
  \hbox{0.1 (yellow), 0.05 (green),}
  \hbox{0.01 (blue).}
  }$}
  \label{ftc2}
 \end{minipage}
 \begin{minipage}{0.5\hsize}
  \begin{center}
   \includegraphics[width=6.0cm]{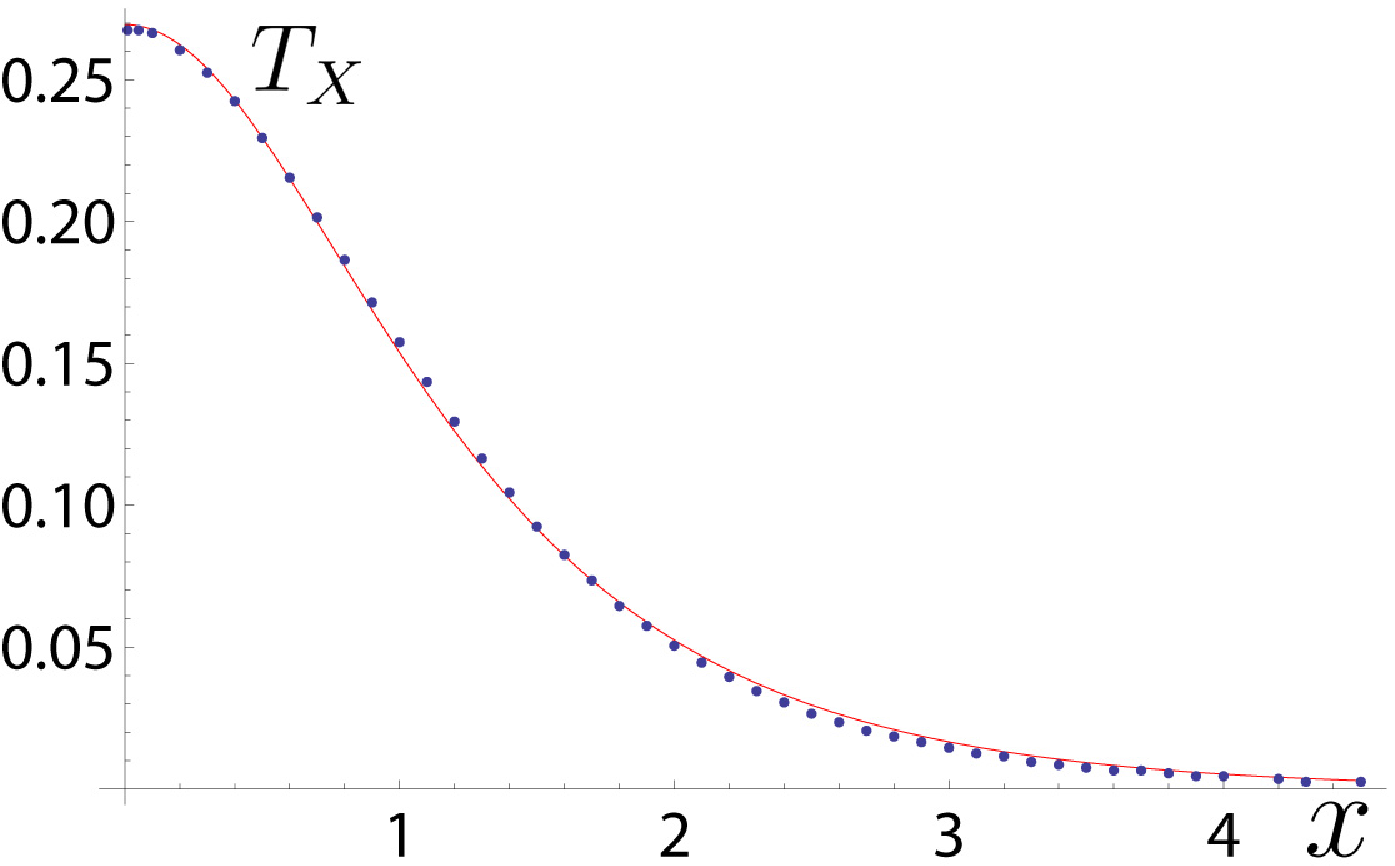}
  \end{center}
  \caption{$\vtop {
  \hbox{Crossover temperature $T_X$}
 \hbox{versus impurity parameter $x$. } 
 \hbox{Estimates of $T_X$ for each $x$ (blue dots). }
  \hbox{Fitting curve: $T_X={a}/{\cosh b x} $ }
  \hbox{with $a=0.269$, $b=1.160$ (red curve). }
  }$}
  \label{fT}
 \end{minipage}
\end{figure}

We define the crossover temperature by the temperature at which 
the specific heat shift $\D c(x, T)$ vanishes. 
We plot the estimates of the crossover temperature $T_X$ 
as a function of the impurity parameter $x$ in Fig. \ref{fT}. 
Furthermore, we find that the following formula gives a good fitting curve to the numerical estimates of the crossover temperature:    
\begin{equation} 
T_{X}=\fr{a}{\cosh b x} \label{eq:TX}
\end{equation}
where the best estimates of parameters  $a$ and $b$ are given 
by $a=0.269$ and $b=1.160$, respectively. 

It is suggested from the impurity susceptibility (\ref{eq:sc-imp}) that the Kondo temperature $T_K$ is given by $T_K \propto 1/\cosh(\pi x/\zeta) $. However, the numerical estimate of parameter $b$ of crossover temperature $T_X$ in (\ref{eq:TX}) is given by almost half the value of $\pi/\zeta = 3$. Thus, the functional form of 
the crossover temperature $T_X$ is the same with the Kondo temperature $T_K$, while the estimate of the parameter $b$ is different.

%
%
\setcounter{equation}{0}
\section{Wilson ratio}\label{Wil}

\subsection{Universality at low temperature for small $|x|$}

Let us recall that we have defined the Wilson ratio by eq. \re{eq:W-ratio}, 
which is given by the ratio of the impurity magnetic susceptibility 
$\chi_{imp}(x)$ to the impurity specific heat $c_{imp}(x, T)$ with some normalization factor. 
We also recall that the impurity susceptibility is given by eq. \re{eq:sc-imp} as 
$$
\chi_{imp}(x) = \fr{4 \z }{\pi(\pi-\z) \sin \z} \cosh\fr{\pi x}{\z} . 
$$
We have shown in \S 4 that when the absolute value of the impurity parameter $x$ is small,    
 the impurity specific heat $c_{imp}(x,T)$ is given by eq. \re{eq:cimp-lowT} for low temperature.  
Thus, by making use of  \re{eq:sc-imp} and \re{eq:cimp-lowT}, 
we evaluate the Wilson ratio \re{eq:W-ratio} in the XXZ impurity model  
for small $x$ at low temperature $T$ by 
\begin{eqnarray}\label{wr}
 r = \fr{\pi^2}{3} \, {\fr{\chi_{imp}} {c_{imp}/T} }
 =\fr{2 \pi}{\pi-\z}.
\end{eqnarray}
We observe that the Wilson ratio $r$ does not depend on the impurity parameter $x$. 
Therefore, the Hamiltonians of the XXZ impurity model with the same value of $\z$ 
but different values of $x$ are classified in the same universality class. 

Furthermore, the Wilson ratio \re{wr} is expressed in terms of the dressed charge ${\cal Z}$ as 
\begin{equation} 
  r = 4 {\cal Z}^2 . \label{wr-dc}
\end{equation}
Here the dressed charge ${\cal Z}$ denotes the dressed charge function 
 ${\cal Z}(\lambda)$ evaluated at the Fermi point: 
${\cal Z}= {\cal Z}(B)$. The dressed charge  ${\cal Z}(\lambda) $ is 
defined by the solution of the integral equation \cite{BIK1986}: 
\begin{equation}
{\cal Z}(\lambda) + \int_{-B}^{B} a_2(\lambda - \mu) {\cal Z}(\mu) d \mu = 1 \, . 
\label{eq:DC}
\end{equation}
It is shown in Ref. \cite{BIK1986} that by applying the Wiener-Hopf method 
to eq. \re{eq:DC} for very small magnetic field $h$ 
 we have  
\begin{equation}
{\cal Z}(B) = \sqrt{ \frac {\pi} {2(\pi - \zeta)}  } \, . 
\end{equation}
Therefore, we obtain relation \re{wr-dc}.  The Wilson ratio of the XXZ impurity model \re{eq:hxxz} 
is expressed in terms of the dressed charge of the XXZ spin chain.

From the CFT viewpoint we give a physical argument for the Wilson ratio expressed with the dressed charge \re{wr-dc} at low temperature for small  $|x|$ in Appendix B.   

%
%
\subsection{Numerical estimates of the Wilson ratio}

\begin{figure}[h]
 \centering
 \includegraphics[width=7cm]{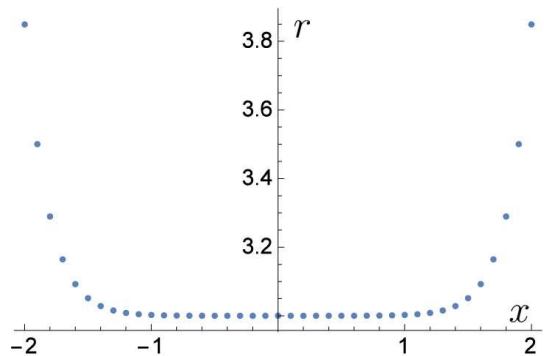}
 \caption{$\vtop { 
\hbox{Wilson ratio $r$ versus impurity parameter $x$ for $\z=\fr{\pi}{3 }$ }
\hbox{through the numerical estimates of $\chi_{imp}(x)$ and $c_{imp}(x, T)$ } 
\hbox{which are at zero temperature with $N=10^4$ and at low } 
\hbox{temperature with $T=10^{-3}$, respectively  (see also,  eq. \re{eq:crossoverT}).}
}$}
 \label{fr}
\end{figure}

The numerical estimates of the Wilson ratio are plotted in  Fig.  \ref{fr} against impurity parameter $x$.  Here we have evaluated the Wilson ratio numerically by making use of the estimates of the impurity susceptibility $\chi_{imp}(x)$ of Fig. \ref{fchi1} in \S 3 and those of the impurity specific heat $c_{imp}(x, T)$ at low temperature with $T=10^{-3}$ of Fig. \ref{fc1} in \S 4.

For small absolute values of impurity parameter $x$ such as $|x| < 1$ the estimate of the Wilson ratio is almost constant with respect to $x$ and given by 3.0. We suggest that the low-temperature impurity effect with small $|x|$ is due to the Kondo effect and the estimate of the Wilson ratio is consistent with eq. \re{wr}. Thus, the universality class is described by the XXZ coupling through the dressed charge as shown in eq. \re{wr-dc}.  

For large absolute values of impurity parameter $x$ such as $|x| > 1$,  however, the estimate of the  Wilson ratio shown in Fig. \ref{fr} increases with respect to $|x|$.  We suggest that it reflects the crossover of the impurity spin becoming a free spin,  
for which the Wilson ratio is given by infinity. 
However, we also suggest that if the numerical estimate of the Wilson ratio 
larger than 3.0 such as for $|x| > 1.4$ is not universal and may depend on the temperature $T$ at which the impurity specific heat  $c_{imp}(x, T)$ is evaluated.  
We expect that the region of $x$ where the estimate of the Wilson ratio is given by $3.0$ 
becomes wider if the temperature $T$ decreases. 

%
%

\vskip 24pt 

\section*{Acknowledgment}
The authors would like to thank F.H.L. Essler and K. Sakai for helpful comments. 
In particular, we thank K. Sakai for useful comments on the Wiener-Hopf method  
and the specific heat of the XXZ spin chain. We also thank S. Okuda for helpful discussion. 
We would also like to thank A. Kl\"{u}mper for helpful comments during the workshop RAQIS '14, Dijon, France.    
The present study is partially supported by Grant-in-Aid for Scientific Research 
No. 24540396.  

%
%
\appendix
\def\thesection{\Alph{section}}

\setcounter{equation}{0}
\setcounter{figure}{0}

%
\section{Expression of  $\xxz(x)$ with spin operators}\label{apph}

We now derive expression \re{eq:hxxz} of the XXZ Hamiltonian with a spin-1/2 impurity which is defined by 
the logarithmic derivative of the transfer matrix as shown in eq. \re{h}. 
Here we express  the Hamiltonian explicitly in terms of  local spin operators 
$S_j^{\pm}$ and $S_j^z$. 
First, we remark that the $R$-matrix at zero rapidity $R_{i,j}(0)$ gives the permutation operator  $P_{i,j}$ acting on the tensor product space $V_j \otimes V_j$:  $R_{i,j}(0)=P_{i,j}$. 
We consider the two factors of  the logarithmic derivative \re{h}, separately. 
\begin{eqnarray}
 \xxz(x)
 &=&\frac{\vp(i\zeta)}2\left[ \tau_{1\cdots N}\left(\frac{i\zeta}{2}\Big|\xi_1,\frac{i\zeta}{2},\cdots,\frac{i\zeta}{2}\right)\right]^{-1} \nn \\
&&\times
 \left[ \dlam\tau_{1\cdots N}\left(\lam\Big|\xi_1,\frac{i\zeta}{2},\cdots,\frac{i\zeta}{2}\right)\Big|_{\lam\to\frac{i\zeta}{2}}\right] \, . \label{eq:logDV} 
\end{eqnarray}
For the first factor of (\ref{eq:logDV}) we have 
\begin{eqnarray}\label{pre}
 \tau_{1\cdots N}\left(\frac{i\zeta}{2}\Big|\xi_1,\frac{i\zeta}{2},\cdots,\frac{i\zeta}{2}\right)
=\Pi^{2\cdots N}R_{N,1}(x) , 
\end{eqnarray}
where $\Pi^{2\cdots N}$ denotes the cyclic permutation operator, which is also expressed as 
the product of permutation operators: $\Pi^{a b c \cdots d}=P_{ad} \cdots P_{ac} P_{ab}$. 
For the second factor of  (\ref{eq:logDV}) we have  
\begin{eqnarray}\label{rest}
 &&\frac{\vp(i\zeta)}2\left[ \dlam\tau_{1\cdots N}\left(\lam\Big|\xi_1,\frac{i\zeta}{2},\cdots,\frac{i\zeta}{2}\right)\Big|_{\lam\to\frac{i\zeta}{2}}\right]
 \nn \\
 &=&\sum_{n=2}^N
\tr0 [P_{0N}\cdots h_{0n} \cdots P_{02}R_{01}(x)]+\tr0 [P_{0N}\cdots P_{02}h_{01}(x)], 
\end{eqnarray}
where symbols $h_{i,j}(\lam)$ denote the derivatives of the $R$-matrices with respect to rapidity $\lambda$   
\begin{equation} 
h_{i,j}(\lam)=\frac{\vp(i\zeta)}2\frac\d{\d\lam}R_{i,j}(\lam).
\end{equation} 

For eq. \re{rest} we consider three cases in the following:

\vspace{10mm}
\noindent 
$\bullet$ For $3 \leq n \leq N$ we have 
\begin{eqnarray}
 \tr0 [P_{0N}\cdots h_{0n} \cdots P_{02}R_{01}(x)]
 =\Pi^{2\cdots n-1} R_{n-1,1}(x) \Pi^{n-1, n+1, \cdots N} h_{n-1,n}. 
\label{eq:hn}
\end{eqnarray}
Applying the inverse of \re{pre} to \re{eq:hn}, we have 
\begin{eqnarray}
 R^{-1}_{N,1}(x)\(\Pi^{2\cdots N}\)^{-1}
 \Pi^{2\cdots n-1} R_{n-1,1}(x) \Pi^{n-1, n+1, \cdots N} h_{n-1,n}
 =\hh_{n-1,n}.
\end{eqnarray}
Here, we used the abbreviation $\check{A}_{i,j}:=P_{i,j}A_{i,j}$.

\vspace{10mm}
\noindent 
$\bullet$ For $n=2$ we have 
\begin{eqnarray}
 \tr0 [P_{0N}\cdots P_{03} h_{02}R_{01}(x)] 
 =\Pi^{3\cdots N} h_{N,2}R_{N,1}(x) . 
\end{eqnarray}
Then, by applying the inverse of \re{pre}, we have
\begin{eqnarray}
 R^{-1}_{N,1}(x)\(\Pi^{2\cdots N}\)^{-1}
 \Pi^{3\cdots N} h_{N,2}R_{N,1}(x)
 =\RR^{-1}_{N,1}(x)\hh_{12}\RR_{N,1}(x).
\end{eqnarray}

\vspace{15mm}
\noindent 
$\bullet$ For $n=1$ we have 

\begin{eqnarray}
 \tr0 [P_{0N}\cdots P_{02}h_{01}(x)]
 = \Pi^{2\cdots N} h_{N,1}(x),
\end{eqnarray}
then
\begin{eqnarray}
 R^{-1}_{N,1}(x)\(\Pi^{2\cdots N}\)^{-1}
\Pi^{2\cdots N} h_{N,1}(x)
=\RR^{-1}_{N,1}(x)\hh_{N,1}(x).
\end{eqnarray}

\vspace{15mm}
Taking the sum of these terms, we have 
\begin{eqnarray} 
 \xxz(x)=\check{R}^{-1}_{N,1}(x)\check{h}_{N,1}
+\check{R}^{-1}_{N,1}(x)\check{h}_{1,2}\check{R}_{N,1}(x)
+\sum_{n=3}^N\check{h}_{n-1,n}. \label{eq:hxxzA} 
\end{eqnarray}
We now express expression \re{eq:hxxzA} in the form of matrices. 
The third term of \re{eq:hxxzA} 
is exactly the same matrix which appears in the XXZ Hamiltonian. Actually, we have 
\begin{eqnarray}
 \sm{n=3}N \hh_{n-1,n}
 &=&\sm{n=3}N \fr12
 \begin{pmatrix}
 0&&&\\
 &-\D&1&\\
 &1&-\D&\\
 &&&0
 \end{pmatrix}_{\!\![n-1,n]} \nn \\
 &=&\sm{n=3}N
 \[ \frac12(S_n^+S_{n+1}^-+S_n^-S_{n+1}^+)+\D(S_n^zS_{n+1}^z-\fr14)\].
\end{eqnarray}
The first term of \re{eq:hxxzA} is given by 
\begin{eqnarray}
 \RR_{N,1}^{-1}(x)\hh_{N,1}(x) 
&=&\fr{c^+c^-}{2}
\begin{pmatrix}
0&&&\\
&-\D&\vp'(x)&\\
&\vp'(x)&-\D&\\
&&&0
\end{pmatrix}_{\!\![N,1]} \nn \\
&=&c^+c^-
\[ \frac{\vp'(x)}2(S_N^+S_{1}^-+S_N^-S_{1}^+)+\D\(S_N^zS_{1}^z-\frac14\)\].
\end{eqnarray}
Here we recall the notation given in eq. \re{bc}. 
The second term of \re{eq:hxxzA} is given by 
\begin{eqnarray}
 &&\RR^{-1}_{N,1}(x)\hh_{12}\RR_{N,1}(x) \nn \\
&=&\fr12
\begin{pmatrix}
(\fr12+S_N^z)+c^-(\fr12-S_N^z)&b^-S_N^-\\
b^-S_N^+&c^-(\fr12+S_N^z)+(\fr12-S_N^z)\\
\end{pmatrix}_{\!\![1]} \nn \\
&&\times
\begin{pmatrix}
-\D(\fr12-S_2^z)&S_2^-\\
S_2^+&-\D(\fr12+S_2^z)\\
\end{pmatrix}_{\!\![1]} \nn \\
&&\times
\begin{pmatrix}
(\fr12+S_N^z)+c^+(\fr12-S_N^z)&b^+S_N^-\\
b^+S_N^+&c^+(\fr12+S_N^z)+(\fr12-S_N^z)\\
\end{pmatrix}_{\!\![1]}.
\end{eqnarray}
By a straightforward calculation, we derive the rest of \re{eq:hxxz}.

\setcounter{equation}{0}

\section{A possible CFT argument for the Wilson ratio} 

Let us explain the Wilson ratio \re{wr-dc} expressed in terms of the dressed charge 
through a physical argument based on the possible CFT.    
We assume that the low-lying excitation spectrum of the 
XXZ spin chain with the spin-1/2 impurity is described by a CFT.

We first derive the ratio of the bulk susceptibility $\chi_s$ 
to the bulk specific heat $C$. 
The magnetic susceptibility $\chi_s^{bulk}$ of the homogeneous XXZ spin chain with no impurity 
is given by \cite{BIK1986}
\begin{equation} 
\chi_s^{bulk} = \frac { 4 {\cal Z}^2} {\pi v_s^{bulk}}  
\end{equation}
where $v_s^{bulk}$ is the group velocity of spin excitation at the Fermi point: 
$v_s= d \epsilon/dk(B) /2 \pi \sigma(B)$ with $\epsilon(k)$ being the dressed energy.   
For the homogeneous XXZ spin chain with no impurity 
the specific heat $C$ at low temperature is given by  
\begin{equation}
C^{bulk} = \frac {\pi }{3 v_s^{bulk}} T . 
\end{equation}
For the bulk susceptibility and the bulk specific heat we thus have 
\begin{equation}
 {\fr {{\pi}^2} 3} \, {\frac{\chi_s^{bulk}} {C^{bulk}/T}}  = 4 {\cal Z}^2  \, . 
\label{eq:CFT-r} 
\end{equation} 

We now extend the calculation for the homogeneous chain\cite{BIK1986} 
to the XXZ chain with the spin-1/2 impurity. We thus have  
\begin{eqnarray} 
\chi_s^{bulk} + {\frac 1 N}  \chi_s^{imp} & = & \frac { 4 {\cal Z}^2} {\pi (v_s^{bulk} +  v_s^{imp}/N )}
\nonumber \\   
& \approx & \frac { 4 {\cal Z}^2} {\pi v_s^{bulk}}  \left( 1 -  {\frac 1 N} {\frac {v_s^{imp}} {v_s^{bulk}}} \right) 
\end{eqnarray} 
where we have assumed the contribution of the impurity in the Fermi velocity, $v_s^{imp}/N$. 
It follows from the  assumption that the low-lying excitation spectrum 
of the XXZ spin chain with the spin-1/2 impurity is given by the CFT of central charge $c=1$ 
that the specific heat is given by 
\begin{eqnarray}
C^{bulk} + {\frac 1 N} C^{imp} & = & \frac {\pi \, T }{3 (v_s^{bulk} + v_s^{imp}/N) } \nonumber \\ 
& \approx  & 
{\frac {\pi \, T }{3 v_s^{bulk}}} \left( 1 -  {\frac 1 N} {\frac {v_s^{imp}} {v_s^{bulk}} }  \right). 
\end{eqnarray} 
Thus, the Wilson ratio for the impurity susceptibility $\chi_s^{imp}$
and the impurity specific heat $C^{imp}$ is given by 
\begin{eqnarray}
{\fr {\pi^2} 3} \, {\fr {\chi_s^{imp}}  {C^{imp}/T}} & =&  
{\fr {\pi^2} 3} \, 
\fr {{ 4 {\cal Z}^2}/{\pi v_s^{bulk}}} 
 {{\pi }/{3 v_s^{bulk}}} \, 
\fr {v_s^{imp}/v_s^{bulk}} 
{v_s^{imp}/v_s^{bulk}} \nonumber \\ 
& = &  4 {\cal Z}^2 \, . \label{wr-CFT}
\end{eqnarray} 
We have thus obtained the Wilson ratio given by eq. \re{wr}. 

For integrable impurity models such as the Anderson and the Kondo models 
 the boundary CFT is applied to characterize the impurity susceptibilities of the models  
\cite{Fujimito-Kawakami-Yang}. However, it seems that it is nontrivial to show that 
some conformal field theory describes the low-lying excitation spectrum 
of the XXZ spin chain with the spin-1/2 impurity. 

We remark that the above argument for the Wilson ratio in eq. \re{wr-CFT} does not depend on the ratio of the impurity group velocity to the bulk one $v_s^{imp}/v_s^{bulk}$. However, it is nontrivial to evaluate the impurity contributions $\chi_s^{imp}$ and $C^{imp}$ so that they    
are consistent with numerical estimates, as we have shown in \S 3 and \S 4 of the present paper.

\end{document}